\documentclass[fleqn,usenatbib]{mnras}
\pdfoutput=1
\usepackage{mathptmx}

\usepackage[T1]{fontenc}
\usepackage{ae,aecompl}

\usepackage{graphicx}
\usepackage{longtable}
\usepackage{amsfonts,amsmath,amssymb}
\usepackage{url}
\usepackage{array}
\newcounter{rowno}
\usepackage[utf8]{inputenc}
\usepackage[english]{babel}

\def\mone{{\rm M}_1}
\def\mtwo{{\rm M}_2}

\def\nsamp{{\rm N}_{\rm samp}}
\def\nmrg{{\rm N}_{\rm mrg}}

\def\dmax{D_{\rm max}}
\def\zmax{z_{\rm max}}
\def\zpeak{z_{\rm peak}}
\def\zf{z_{\rm f}}
\def\tf{t_{\rm f}}
\def\zd{z_{\rm D}}
\def\tevnt{t_{\rm event}}
\def\zevnt{z_{\rm event}}

\def\tld{t_{\rm lD}}
\def\tobs{t_{\rm obs}}
\def\delobs{\Delta t_{\rm obs}}
\def\delage{\Delta t_{\rm age}}

\def\rate{\mathcal R}
\def\rpess{{\mathcal R}_{-}}
\def\rz{{\mathcal R}^{\prime}}
\def\recc{{\mathcal R}_{\rm ecc}}
\def\emx{e_{\rm m}}
\def\rhogc{\rho_{\rm GC}}
\def\clmf{\phi_{\rm CLMF}}
\def\sfh{\Phi_{\rm SFH}}
\def\rmort{{\rm R}_{\rm mort}}

\def\peryg{{\rm~yr}^{-1}{\rm Gpc}^{-3}}

\def\mgclow{M_{\rm GC,low}}
\def\mgchigh{M_{\rm GC,high}}
\def\mcllow{M_{\rm cl,low}}
\def\mclhigh{M_{\rm cl,high}}
\def\permv{{\rm~Mpc}^{-3}}
\def\mcrit{{\rm M}_{\rm crit}}

\newcommand{\Ms}{\ensuremath{{\rm M}_{\odot}}}
\newcommand{\Zs}{\ensuremath{{\rm Z}_{\odot}}}
\newcommand{\eg}{{\it e.g.}}

\newcommand{\ie}{{\it i.e.}}

\newcommand{\beq}{\begin{equation}}
\newcommand{\eeq}{\end{equation}}

\newcommand{\kmps}{\ensuremath{{\rm~km~s}^{-1}}}

\newcommand{\thub}{\ensuremath{t_{\rm Hubble}}}

\newcommand{\mcl}{\ensuremath{M_{\rm cl}}}
\newcommand{\rh}{\ensuremath{r_{\rm h}}}

\newcommand{\tmrg}{\ensuremath{t_{\rm mrg}}}

\newcommand{\nmrgin}{\ensuremath{N_{\rm mrg,in}}}
\newcommand{\nmrgout}{\ensuremath{N_{\rm mrg,out}}}

\newcommand{\nbseven}{{\tt NBODY7}}

\newcommand{\bse}{{\tt BSE}}

\newcommand{\archain}{{\tt ARCHAIN}}

\newcommand{\tevol}{\ensuremath{T_{\rm evol}}}

\newcommand{\fbin}{\ensuremath{f_{\rm bin}}}
\newcommand{\nbin}{\ensuremath{N_{\rm bin}}}

\newcommand{\fmrg}{\ensuremath{f_{\rm mrg}}}
\newcommand{\ftz}{\ensuremath{f_{\rm TZ}}}

\title[Stellar-mass black holes in young clusters V]
{Stellar-mass black holes in young massive and open stellar
clusters V: comparisons with LIGO-Virgo merger rate densities}

\author[S. Banerjee]{
Sambaran Banerjee$^{1,2}$\thanks{E-mail: sambaran@astro.uni-bonn.de (SB)}
\\
$^{1}$Helmholtz-Instituts f\"ur Strahlen- und Kernphysik (HISKP),
Nussallee 14-16, D-53115 Bonn, Germany\\
$^{2}$Argelander-Institut f\"ur Astronomie (AIfA),
Auf dem H\"ugel 71, D-53121, Bonn, Germany
}

\pubyear{2020}

\begin{document}
\label{firstpage}
\pagerange{\pageref{firstpage}--\pageref{lastpage}} 
\maketitle

\begin{abstract}
I study the contribution of young massive star clusters (YMCs) and open star clusters (OCs)
to the present day, intrinsic merger rate density of dynamically-assembled binary black holes (BBHs).
The BBH merger event rate is estimated based on a set of state-of-the-art evolutionary models
of star clusters, as presented in \citet{Banerjee_2020c}.
The merger-event rates are obtained by constructing a cluster population of the Universe,
out of the models, taking into account mass distribution of clusters and
cosmic star formation and enrichment histories, as per observations. The model BBH merger rate density
ranges from a pessimistic to a reference value of $0.5\peryg-37.9\peryg$, for a
LIGO-Virgo-like detector horizon. The reference
rate well accommodates the BBH merger rate densities estimated from GWTC-1 and GWTC-2
merger-event catalogues.
The computed models also yield differential BBH merger rate densities that
agree reasonably with those from GWTC-1 and, as well, with the much more constrained
ones from GWTC-2. These results suggest that dynamical interactions in YMCs and OCs can, in principle,
alone explain the BBH merger rate density and its dependence on the merging-binary properties, as inferred
from to-date gravitational-wave (GW) events.
The cosmic merger rate density evolution also agrees with GWTC-2.
The models predict 
a rate of $\approx5\peryg$ for eccentric LIGO-Virgo mergers from YMCs and OCs.
The improving constraints on BBH merger rate density with mounting GW events 
will help constraining scenarios of star cluster formation across cosmic time
and as well the relative contributions of the various compact binary merger channels.
\end{abstract}


\section{Introduction}\label{intro}

Until 2019, the LIGO-Virgo collaboration (hereafter LVC),
in their first gravitational wave transient catalogue \citep[][hereafter GWTC-1]{Abbott_GWTC1},
has published
11 compact binary merger events from their first and second observing
runs (hereafter O1 and O2, respectively) with the ground-based
interferometric gravitational wave (hereafter GW) detectors
LIGO \citep{Asai_2015} and Virgo \citep{Acernese_2014}. In 2020,
the LIGO-Virgo-KAGRA (hereafter LVK) collaboration has announced,
in their second gravitational wave transient catalogue \citep[][hereafter GWTC-2]{Abbott_GWTC2},
39 additional candidates of compact binary coalescence events
from the first half, `O3a', of their recently concluded third observing run (hereafter O3).
Based on the parameter estimations of these events, the vast majority of 
them has been designated as binary black hole (hereafter BBH) mergers with
component masses ranging through $\approx5\Ms-90\Ms$ \citep{Abbott_GWTC2}.
The rest comprise likely candidates of binary neutron star (hereafter BNS) mergers
and neutron star-black hole binary mergers.

Even prior to the publication of GWTC-1, the handful of then known GW events had
already triggered wide debates regarding possible origins of such merging compact binaries
and of the masses of the black holes (hereafter BH) and
neutron stars (hereafter NS) they are made of. The issues remain as open until now.
In these regards, the large jump in the number of events from GWTC-1 to GWTC-2
\citep[consistently with the improved detector sensitivity and the observing time during O3a, see][]{Abbott_GWTC2}
is particularly enlightening: apart from providing us with a
wider variety and, as well, highly atypical GW events
\citep[\eg,][]{LSC_GW190412,Abbott_GW190814,Abbott_GW190521},
the constraints on the rate of such merger events in the Universe is
significantly improved in GWTC-2. Event rate is among the key
aspects that would help to understand the relative contributions
of the various astrophysical channels (or scenarios) for
general-relativistic (hereafter GR) inspiral and mergers of compact binaries.
Such channels can be divided into two main categories
\citep{Benacquista_2006,Benacquista_2013,Mandel_2017,Mapelli_2018}, namely,
(a), evolution of isolated massive stellar binaries
\citep[\eg,][]{Dominik_2012,Belczynski_2016,Belczynski_2016b,Marchant_2016,Stevenson_2017,Giacobbo_2018,Spera_2019,Rastello_2020,Santoliquido_2020,Belczynski_2020,Bavera_2020}
and, (b), dynamical interactions among stellar remnants in various dense-stellar and dynamically-active environments or systems
such as globular clusters (hereafter GC) \citep[\eg,][]{Breen_2013,Morscher_2013,Rodriguez_2015,Askar_2016,Chatterjee_2017a,Askar_2018,Fragione_2018b,Antonini_2020,Kremer_2020},
nuclear clusters \citep[\eg,][]{Antonini_2016,Hoang_2017,ArcaSedda_2020,ArcaSedda_2020b,Mapelli_2020b},
young massive clusters (hereafter YMC), open clusters (hereafter OC)
\citep[\eg,][]{Banerjee_2010,Ziosi_2014,Mapelli_2016,Banerjee_2017,Rastello_2019,DiCarlo_2019,Kumamoto_2019,Banerjee_2020c},
field hierarchical systems \citep[\eg,][]{Katz_2011,Lithwick_2011,Antonini_2017,Silsbee_2017,Fragione_2019b,Fragione_2020}, and stellar-remnant BHs trapped in gas disks in active galactic nuclei \citep[\eg,][]{McKernan_2018,Secunda_2019}.

Dynamical interactions among stellar-remnant (or stellar-mass) BHs
in YMCs and OCs has recently drawn high interest due to the channel's natural ability to
produce the unusual, highly mass-asymmetric \citep[\eg, GW190412 and GW190814;][]{LSC_GW190412,Abbott_GW190814}
or very massive \citep[\eg, GW190521;][]{Abbott_GW190521} BBH mergers
\citep{DiCarlo_2020,Banerjee_2020c}. Low and moderate mass young clusters, owing to their
relatively short two-body relaxation times \citep{Spitzer_1987,Heggie_2003} and spatial ambience, can produce
BBH mergers at rates comparable to or exceeding those from GCs and
isolated binary evolution \citep{Banerjee_2017b,Kumamoto_2020,DiCarlo_2020,Santoliquido_2020}.
BBH merger events and rates apart, interest in younger evolutionary phases of all categories of stars clusters
would grow naturally with increasing visibility horizons \citep{Chen_2017b} of the forthcoming upgrades
of the current GW detectors (\eg, the LIGO A+ upgrade) and future GW detectors
(\eg, Voyager, Einstein Telescope, Cosmic Explorer; \citealt{Reitze_2019}).
With increasing look back time of the GW sources, one essentially rewinds to
younger versions of the clusters, \ie, accesses mergers of shorter delay times. 

In this work, the set of state-of-the-art N-body evolutionary models
of star clusters, as described in \citet{Banerjee_2020c}, is
utilized to estimate the contribution of dynamical interactions, in intermediate mass and massive
YMCs and OCs, to the present-day BBH merger rate density.
Sec.~\ref{nbcomp} summarizes the computed star cluster models.
Sec.~\ref{ratecalc} describes the method used to evaluate the
present-day, intrinsic BBH merger rate density and the
corresponding differential merger rate densities (w.r.t.
the merging binary's primary mass, mass ratio, and eccentricity),
based on the computed model set and observationally-derived
cluster population properties.
Sec.~\ref{sec.diffr} presents the differential BBH
merger rate densities as estimated from the computed models.
Sec.~\ref{sec.rz} explores how the model BBH merger rate density
depends on the GW detector's horizon and event redshifts.
Sec.~\ref{sec.diffr} and Sec.~\ref{sec.rz} also make detailed comparisons
with the BBH merger rate densities and BBH differential merger rate
densities obtained from GWTC-1 and GWTC-2.
The results, their various uncertainties, and caveats in the present
approach are further discussed in Sec.~\ref{discuss}.
Sec.~\ref{summary} summarizes the results and discusses potential next steps.

\section{Method}\label{method}

Below, the evolutionary star cluster models and the method
for calculating the present-day merger rate density are described.

\subsection{Direct N-body star cluster-evolutionary models
with up-to-date remnant formation and post-Newtonian dynamics}\label{nbcomp}

In this work, the 65 N-body evolutionary models of star clusters,
as described in \citet[][hereafter Paper II]{Banerjee_2020c}, are utilized.
These computations and the model ingredients are described in detail in Paper II
and further discussions are provided in \citet{Banerjee_2020d}.
Therefore, only a summary of these computations is provided in this paper
as follows.

The model clusters,
initially, have a Plummer density profile \citep{Plummer_1911},
are virialized \citep{Spitzer_1987,Heggie_2003}, and are unsegregated
(\ie, have no radial dependence of stellar mass distribution). They, initially,
have masses of $10^4\Ms\leq\mcl\leq10^5\Ms$ and half-mass radii of 
$1{\rm~pc}\leq\rh\leq3{\rm~pc}$. They range over 
$0.0001\leq Z \leq0.02$ in metallicity and orbit in a solar-neighborhood-like
external galactic field. The initial models are made of zero-age-main-sequence
(hereafter ZAMS) stars with masses of $0.08\Ms\leq m_\ast\leq150.0\Ms$
which are distributed according
to the standard initial mass function (hereafter IMF; \citealt{Kroupa_2001}).
About half of these models
have a primordial-binary population (overall initial binary fraction
$\fbin\approx5$\% or 10\%) where all the O-type stars (\ie, stars
with ZAMS mass down to $\mcrit=16\Ms$) are initially paired among themselves
according to an observationally-motivated distribution of massive-star binaries
\citep{Sana_2011,Sana_2013,Moe_2017}. Such cluster
parameters and stellar compositions are consistent with those
observed in `fully-formed', (near-)spherical, (near-)gas-free YMCs and
medium-mass OCs \citep{PortegiesZwart_2010,Banerjee_2017c,Banerjee_2018b}
that continue to form and dissolve in the Milky Way and other Local-Group galaxies.

These model clusters are evolved using $\nbseven$,
a state-of-the-art post-Newtonian (hereafter PN) direct N-body integrator
\citep{Aarseth_2003,Aarseth_2012,Nitadori_2012}, that couples with the
semi-analytical stellar and binary-evolutionary model
$\bse$ \citep{Hurley_2000,Hurley_2002}. The integrated $\bse$ is made
up to date, in regards to prescriptions of stellar wind mass loss
and formation of NSs and BHs,
as described in \citet[][hereafter Paper I]{Banerjee_2020}.
In particular, the NSs and BHs form according to
the `rapid' and `delayed' SN models of \citet{Fryer_2012} and
pulsation pair-instability SN (PPSN) and pair-instability SN (PSN) 
models of \citet{Belczynski_2016a}. The NSs and BHs
receive natal kicks based on SN fallback onto them,
as in \cite{Belczynski_2008}. Such fallback slows down
the remnants, causing BHs of $\gtrsim10\Ms$ (Paper I)
to retain in the clusters right after their birth. Furthermore,
NSs formed via electron-capture SN \citep{Podsiadlowski_2004}
also receive small natal kicks and retain in the clusters.

In $\nbseven$, the PN treatment is handled by the $\archain$ algorithm
\citep{Mikkola_1999,Mikkola_2008}. Such a PN treatment allows for
GR evolution of the innermost NS- and/or BH-containing binary
of an in-cluster (\ie, gravitationally bound to the cluster)
triple or higher order compact subsystem, in tandem with the Newtonian-dynamical
evolution of the subsystem (Kozai-Lidov oscillation or
chaotic three-body interaction), potentially leading
to the binary's (in-cluster) GR in-spiral and merger. The treatment also
undertakes the GR evolution of in-cluster NS/BH-containing binaries that
are not a part of a higher-order subsystem. As discussed in Paper II
(see also the references therein), the moderate density and velocity dispersion
in the model clusters make them efficient factories of dynamically
assembling PN subsystems, particularly, those comprising BHs.
As also discussed in Paper II \citep[see also][]{Anagnostou_2020},
the vast majority of the GR mergers
from these computed model clusters are in-cluster BBH mergers.
As also demonstrated therein \citep[see also][]{Banerjee_2020d},
the final in-spiralling phases of such merging BBHs
sweep through the LISA and deci-Hertz GW frequency bands
before merging in the LVK band.

\begin{table*}
	\caption{A summary of the `mock detection experiments' performed in this work (Sec.~\ref{ratecalc}).
	The columns
	from left to right are as follows: Col.~1: the total number of clusters in the sample comprising
	the Model Universe, $\nsamp$, Col.~2: the total number of merger events, $\nmrg$, from this sample
	at the present cosmic epoch, within a time $2\delobs$ (Col.~3) around the current age
	of the Universe, Col.~4: the instrument visibility boundary, $\zmax$, for average source inclination,
	Col.~5(6): the inferred reference (pessimistic) present-day, intrinsic merger rate density
	from the Model Universe (see Sec.~\ref{ratecalc}, Eqn.~\ref{eq:rate}).
	}
\label{tab_rates}
\centering
\begin{tabular}{cccccc}
\hline
$\nsamp$      & $\nmrg$ & $\delobs/[{\rm Gyr}]$ & $\zmax$ & $\rate/[\peryg]$ & $\rpess/[\peryg]$ \\ 
\hline
$5\times10^5$ & 16407     & 0.15                  & 1.0     &   37.9           &   0.51            \\
$5\times10^5$ & 26795     & 0.15                  & 2.0     &   61.9           &   0.84            \\
$5\times10^5$ & 27775     & 0.15                  & 3.5     &   64.1           &   0.87            \\
$5\times10^5$ & 24493     & 0.15                  & 5.0     &   56.6           &   0.76            \\
$5\times10^5$ & 19121     & 0.15                  & 7.5     &   44.1           &   0.60            \\
$5\times10^5$ & 16335     & 0.15                  & 10.0    &   37.7           &   0.51            \\
$5\times10^5$ & 13006$^a$ & 0.15                  & 1.0     &   132.2          &   0.99            \\
$5\times10^5$ & 33741$^b$ & 0.15                  & 1.0     &   19.5           &   0.26            \\
$5\times10^5$ & 15162$^c$ & 0.15                  & 1.0     &   35.0           &   0.47            \\
\hline
{\scriptsize $^a$ $\clmf(\mcl)\propto \mcl^{-2.5}$} & & & & & \\
{\scriptsize $^b$ $\mcllow=5\times10^4\Ms$} & & & & & \\ 
{\scriptsize $^c$ models of Table~\ref{tab_nbrun}} & & & & &  
\end{tabular}
\end{table*}

\subsection{Calculation of intrinsic merger rate density
from computed model clusters}\label{ratecalc}

To obtain the intrinsic merger rate density at the present cosmic epoch,
`mock detection experiments' are performed in a `Model Universe'
that is constructed out of the computed model clusters of Paper II.
An ideal GW detector (a detector with zero noise floor)
is considered that can detect GW arriving from all GR mergers within a comoving
volume whose boundary is at a redshift $\zmax$. $\zmax$ represents an
(artificial) horizon, for average source inclination, for a realistic
GW detector like LVK. For LVC O1/O2 observing runs,
$\zmax\approx1$ \citep{Chen_2017b,Abbott_GWTC1_prop}. However, since an ideal GW detector
is considered here, $\zmax$ will be
varied to also address future, ground-based GW detectors of $>$ Hz frequency band
(\eg, LIGO A+ upgrade, Einstein Telescope, Cosmic Explorer).
If the GR compact-binary merger rate density (per unit comoving volume), $\rz$,
in the Universe has an inherent dependence on the merger event
redshift, $\zevnt$, then the present day, intrinsic merger rate density, $\rate$, would
depend on the detector horizon, $\zmax$ \citep{Abadie_2010,Abbott_GWTC1_prop}.
Conversely, if $\rz$ is independent of $\zevnt$, then $\rate$ will also be independent
of $\zmax$. An inherent $\rz(\zevnt)$ dependence would arise from a
non-uniform distribution of delay time
(time of merger since the birth of the parent stellar population or host system),
depending on the channel(s) responsible for GR mergers in the Universe
\citep{Benacquista_2006,Benacquista_2013}, and as well
due to the cosmic variation of star formation rate. Hereafter, for brevity, the present-day,
(differential), intrinsic merger rate density, $\rate$, will simply be referred to as
(differential) merger rate density. The inherent redshift dependence of the merger
rate density, $\rz(\zevnt)$, will, hereafter, be referred to as the
cosmic merger rate density.

From the computed grid (see Table C1
of Paper II), model clusters are randomly chosen with initial masses, within
$2\times10^4\Ms\leq\mcl\leq10^5\Ms$\footnote{Since $\mcl=10^4\Ms$ clusters
are sparse in the model set, clusters of $\mcl\geq2\times10^4\Ms$ are
considered in this work.},
according to a power law distribution
of index $-2$ (\ie, $\clmf(\mcl)\propto \mcl^{\alpha}$; $\alpha=-2$)
as observations of young clusters in the Milky Way and
nearby galaxies suggest \citep{Gieles_2006b,Larsen_2009,PortegiesZwart_2010,Bastian_2012}.
Their initial sizes are chosen uniformly between $1{\rm~pc}\leq\rh\leq3{\rm~pc}$.
Each selected cluster is then assigned a formation redshift, $\zf$,
that corresponds to an age, $\tf$, of the Universe, 
according to the probability distribution given
by the cosmic star formation history \citep[][hereafter SFH]{Madau_2014}
\beq
\sfh(\zf) = 0.015\frac{(1+\zf)^{2.7}}{1+[(1+\zf)/2.9]^{5.6}} \Ms{\rm~yr}^{-1}{\rm~Mpc}^{-3}.
\label{eq:mdsfr}
\eeq

Since the masses of the stellar-remnant BHs (and hence of the merging BBHs) depend
on their parent cluster's metallicity, $Z$ (see Paper I and references therein),
the mass dependence of the differential
merger rate density would depend on the $Z$-distribution of star clusters in the Universe
and the distribution's redshift ($z$) dependence. In this work, the observationally-derived
$Z$-spread and its $z$-dependence over $0\leq z\leq10$, as obtained by \citet{Chruslinska_2019}
(based on their `moderate-$Z$' sample), is adopted \citep[see also][]{Chruslinska_2020}.
The metallicity of a model cluster
of $\mcl$, $\rh$, $\zf$ ($\tf$) is selected from the model grid based on this observationally-derived $Z-z$
distribution\footnote{In practice, a $100\times1000$ $Z-z$
matrix is generated using their publicly-available \texttt{moderate\_FOH\_z\_dM.dat}
and the corresponding {\tt Python} script. The \citet{Anders_1989} solar metallicity scaling
is adopted to covert their O/H-metallicity to Fe-metallicity, $Z$. A $Z$ is randomly
picked from the $Z-z$ lookup table, for the tabulated $z$ that is closest to $\zf$.
The model with metallicity closest to this $Z$ is then selected. See also \citet{Kumamoto_2020}.}.

In this way, the comoving volume within the detector horizon, $\zmax$, is uniformly populated
with a sample cluster population of size $\nsamp=5\times10^5$. A GR merger occurs
from a cluster after a delay time, $\tmrg$, from the cluster's formation
when the age of the Universe is $\tevnt$, \ie,
\beq
\tevnt = \tf + \tmrg. 
\label{eq:tevnt}
\eeq
If the light travel time from the cluster's comoving (or Hubble) distance, $D$, is $\tld$, then
the age of the Universe is
\beq
\tobs = \tevnt + \tld
\label{eq:tobs}
\eeq
when the (redshifted) GW signal from the merger event arrives the detector.
The GW signal is considered `present-day' (or `recent' or `in the present epoch') if
\beq
\thub-\delobs \leq \tobs \leq \thub+\delobs
\label{eq:delobs}
\eeq
where $\thub$ is the current age of the Universe (the Hubble time) and $\delobs$ is taken
to be $\delobs=0.15{\rm~Gyr}$. $\delobs$ serves as an uncertainty in the cluster formation
epoch; with the above choice it is well within the typical epoch uncertainties in the
observed SFH data \citep{Madau_2014}. In this work, SFH within $z\leq10$ is considered
to be consistent with the adopted cosmic metallicity evolution (see above).

If $\nmrg$ present-day mergers are obtained from $\nsamp$ clusters, then the number
of mergers per cluster, $\nmrg/\nsamp$, can be scaled to infer the merger rate density
from the Model Universe. The scaling factor is simply the total number of clusters 
formed, from $z=10$ until $z=0$ and with masses $2\times10^4\Ms\leq\mcl\leq10^5\Ms$
(consistently with the initial mass range of the model clusters and
the cosmic metallicity evolution with which the sample Model-Universe cluster population
is constructed),
within the comoving volume enclosed by $\zmax$ (corresponding to the comoving distance $\dmax$),
divided by $2\delobs$. Hence, the merger rate density is given by 

\begin{equation}
\begin{aligned}
\rate = & \left(\frac{\nmrg}{\nsamp}\right)\left(\frac{1}{2\delobs}\right)\\
	& \left[\frac{\int_{\mcllow}^{\mclhigh}\clmf(\mcl)d\mcl}
	       {\int_{\mgclow}^{\mgchigh}\clmf(\mcl)d\mcl}\right]
\left[\frac{\int_{10}^{0}\sfh(\zf)d\zf}{\int_6^3\sfh(\zf)d\zf}\right]\rhogc\rmort
\end{aligned}
\label{eq:rate}
\end{equation}

In Eqn.~\ref{eq:rate}, the scaling is implemented with respect to the observed number density of GCs per unit
comoving volume, $\rhogc$. In this equation, the first ratio of integrals in the r.h.s. 
is the `boost factor' due to the difference between the birth mass range of YMCs and OCs
considered here, $[\mcllow,\mclhigh]$, and the birth mass range of GCs, $[\mgclow,\mgchigh]$.
Given the steep power-law nature of $\clmf(\mcl)$ (see above), $\rate$ would be sensitive
to these mass ranges and especially on $\mcllow$ and $\mgclow$ which, therefore, deserve
careful considerations.
To be consistent with the sample cluster population with which the Model Universe is constructed,  
$[\mcllow,\mclhigh]=[2\times10^4\Ms,1\times10^5\Ms]$ (but see Sec.~\ref{discuss}).
The birth mass range of GCs is chosen based on the initial masses of the GC models
in the CMC Cluster Catalog described in \citet{Kremer_2020}, which reproduce the observed
present-day mass range of Milky Way GCs by evolving
(using the Monte Carlo approach; \citealt{Henon_1971,Fregeau_2007,Hypki_2013}) from such initial
masses. In \citet{Kremer_2020}, the typical present-day GC masses are obtained 
from initial masses within $[\mgclow,\mgchigh]=[5\times10^5\Ms,1\times10^6\Ms]$ which GC initial
mass range is adopted here for evaluating the `reference' value of $\rate$. The entire (present-day)
Milky Way GC mass range (\ie, including the least and most massive GCs) is obtained
from $[\mgclow,\mgchigh]=[1\times10^5\Ms,2\times10^6\Ms]$ which GC initial mass range
is used for obtaining a `pessimistic' $\rate$ (hereafter $\rpess$). 

The second ratio of integrals in the r.h.s. of Eqn.~\ref{eq:rate} is the boost factor
due to the difference in SFH between the progenitors of YMCs/OCs and what we call
GCs \citep{Harris_1996}.
Star clusters like YMCs (some authors refer to young clusters of $\gtrsim10^5\Ms$
as `super star clusters', conceived to be the young progenitors of present-day GCs)
continue to form throughout the cosmic star formation history
and evolve to become old OCs or GCs or dissolve by the present cosmic epoch. Therefore,
in the numerator, $\zf$ is considered over $[0.0,10.0]$, the limit at $\zf=10.0$ being
due to that in the metallicity evolution history \citep{Chruslinska_2019} used
to construct the sample cluster population (see above). This is also why $\zmax\leq10.0$ in this work
(see Fig.~\ref{fig:rzmax}). The GCs, on the other hand (by definition), are all old objects which
have formed over $3.0\lesssim\zf\lesssim6.0$ \citep{ElBadry_2019b}, setting the $\zf$ limits in the
denominator.

The spatial number density of GCs (per unit comoving volume), $\rhogc$, in
Eqn.~\ref{eq:rate} is taken to be the observationally-determined value,
$\rhogc=8.4h^3\permv$, as in \citet{PortegiesZwart_2000}, for the reference $\rate$.
With the dimensionless
Hubble constant $h\equiv H_0/[100\kmps{\rm~Mpc}^{-1}]=0.674$ \citep{Planck_2018},
$\rhogc=2.57\permv$. For obtaining $\rpess$, $\rhogc=0.33\permv$ is taken which
is a lower limit of GC spatial density as estimated in \citet{Rodriguez_2015}.   

The factor $\rmort$ in Eqn.~\ref{eq:rate} is the `mortality ratio' that absorbs
any inherent inefficiency, relative to GC progenitors, to become a typical gas-free
young cluster from a gas-embedded, proto-cluster phase.
Effects relating to star (cluster) formation mechanisms and environment
\citep[\eg,][]{Banerjee_2018b,Kruijssen_2019}
would determine the success of assembling a gas-free, parsec-scale, young
cluster of the kind we typically observe (and as taken as initial conditions of
the model clusters). The cluster formation efficiency can
depend on the gas-free initial mass, $\mcl$, and formation epoch, $\zf$,
which are, still, largely open questions \citep[\eg,][]{Renaud_2018,Krumholz_2019}.
In this work, for simplicity, $\rmort=1$ is assumed implying that,
beyond $\mcl\gtrsim10^4\Ms$ (as in the models here), the cluster formation efficiency
is assumed to be independent of $\mcl$. For example, direct N-body models suggest
that embedded clusters of $\gtrsim10^4\Ms$ are resilient to the violent relaxation
phase induced by rapid residual gas expulsion from proto-clusters
\citep[\eg,][]{Brinkmann_2017,Shukirgaliyev_2017}.

If the normalized present-day distribution of a quantity $X$,
that is measurable from the detected
merger-event GW signals (\eg, the merging compact binary's primary mass, mass ratio,
eccentricity), is $\psi(X)$ over the range $[X1,X2]$ (\ie, $\int_{X1}^{X2}\psi(X)dX=1$)
then the differential merger rate density w.r.t. $X$ is obtained by
\beq
\frac{d\rate}{dX}(X)=\rate\psi(X).
\label{eq:diffrate}
\eeq
(Hence, $\int_{X1}^{X2}[d\rate/dX(X)]dX=\rate$.)
Here, $\rate$ is determined from Eqn.~\ref{eq:rate}.
In the present mock detection experiment, $[X1,X2]$ is divided into $N_b$ equal-sized
bins of width $\Delta X$. If the number of events detected within the $i$-th bin
around $X_i$ is $\Delta N_{X,i}$ ($\sum_{i=1}^{N_b}\Delta N_{X,i}=\nmrg$) then
\beq
\psi(X_i)\approx\frac{\Delta N_{X,i}}{\Delta X\nmrg}.
\label{eq:diffrate2}
\eeq
(Hence, $\sum_{i=1}^{N_b}\rate\psi(X_i)\Delta X=\rate$.)

In this study, redshift, comoving distance, and light travel time are interrelated
\citep[based on a lookup table;][]{Wright_2006}
according to the $\Lambda$CDM cosmological framework \citep{Peebles_1993,Narlikar_2002}.
The cosmological constants from the latest Planck results
($H_0=67.4\kmps{\rm~Mpc}^{-1}$, $\Omega_{\rm~m}=0.315$, and flat Universe
for which $\thub=13.79{\rm~Gyr}$; \citealt{Planck_2018}) are applied.

In summary, the reference (differential) merger rate density, $\rate$ ($d\rate/dX$)
\footnote{In a few instances in this text, the symbol $\rate$ is also used to
denote the merger rate density from Eqn.~\ref{eq:rate} in general, as
evident from the context. Due to only a handful of such occurrences in this text,
no new symbol is invoked.},
is obtained from Eqn.~\ref{eq:rate}
(and Eqns.~\ref{eq:diffrate}, \ref{eq:diffrate2}) with $[\mgclow,\mgchigh]=[5\times10^5\Ms,1\times10^6\Ms]$
and $\rhogc=2.57\permv$. A (double-)pessimistic (differential) rate, $\rpess$ ($d\rpess/dX$),
is obtained with
$[\mgclow,\mgchigh]=[1\times10^5\Ms,2\times10^6\Ms]$ and $\rhogc=0.33\permv$
in Eqn.~\ref{eq:rate} (and Eqns.~\ref{eq:diffrate}, \ref{eq:diffrate2}).
For both evaluations, $[\mcllow,\mclhigh]=[2\times10^4\Ms,1\times10^5\Ms]$ (but see Sec.~\ref{discuss}).
$\delobs=0.15{\rm~Gyr}$ is used for all the mock detection experiments
\footnote{The unit of $\rate$ ($d\rate/dX$) is then 
${\rm Gyr}^{-1}\permv([X]^{-1})=\peryg([X]^{-1})$.}. $N_b=10$ or 20 is used.

\begin{figure*}
\includegraphics[width=8.7cm,angle=0]{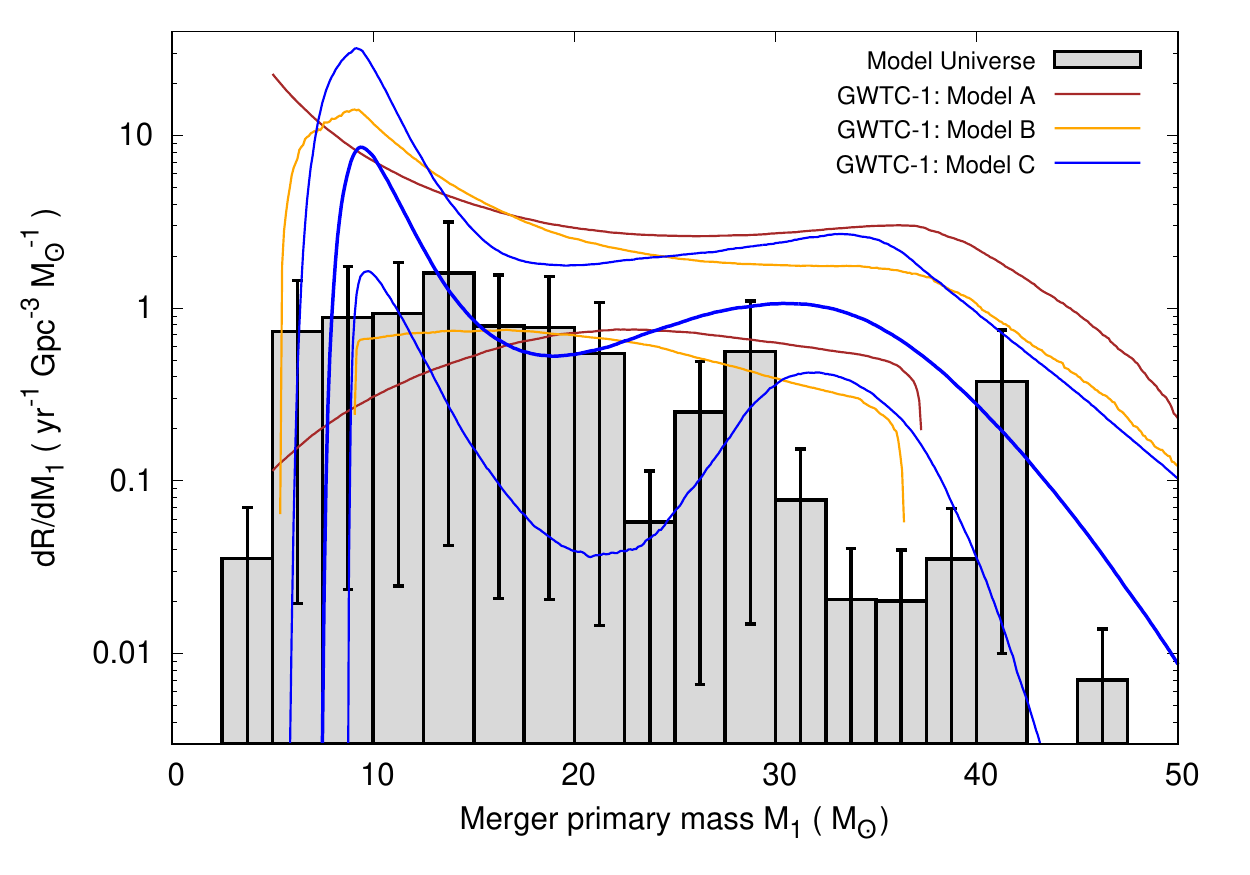}
\includegraphics[width=8.7cm,angle=0]{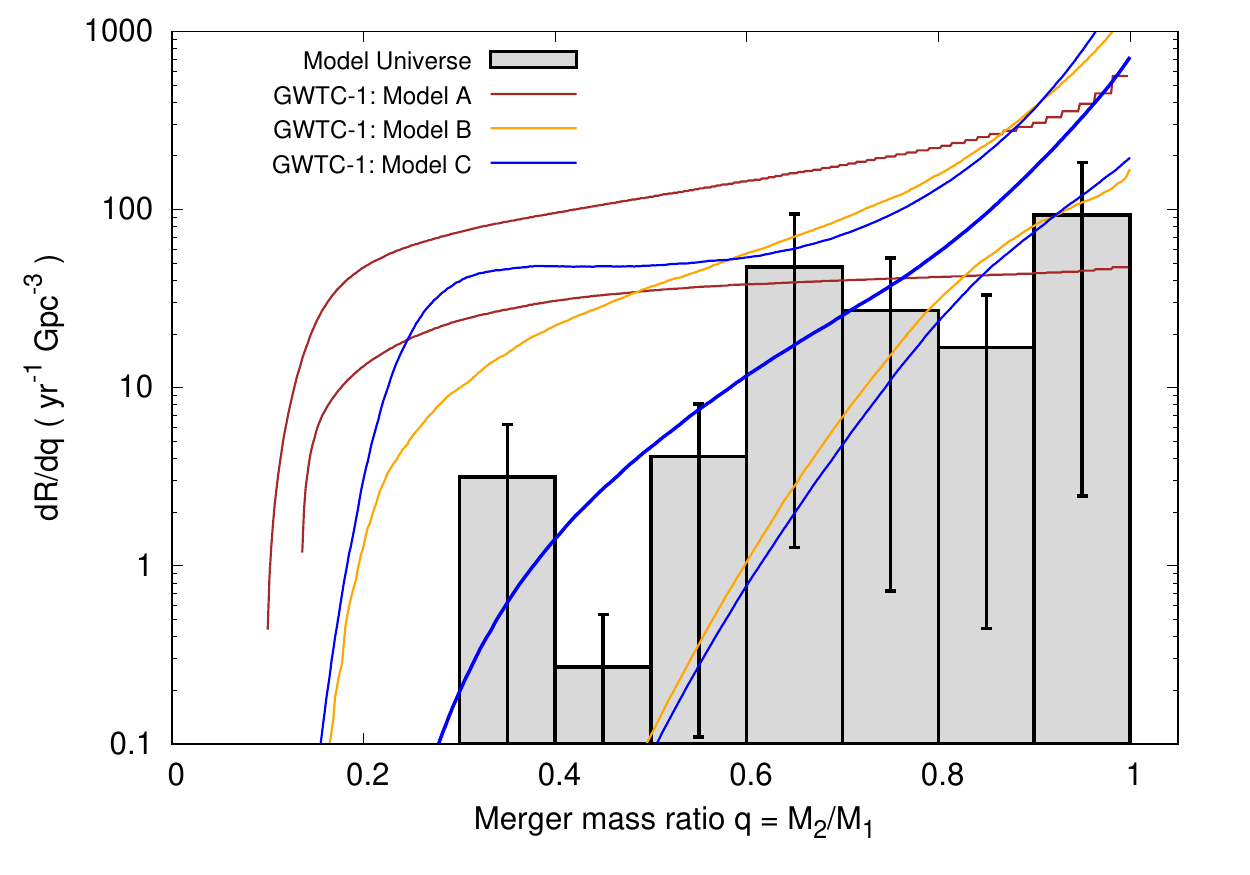}\\
\includegraphics[width=8.7cm,angle=0]{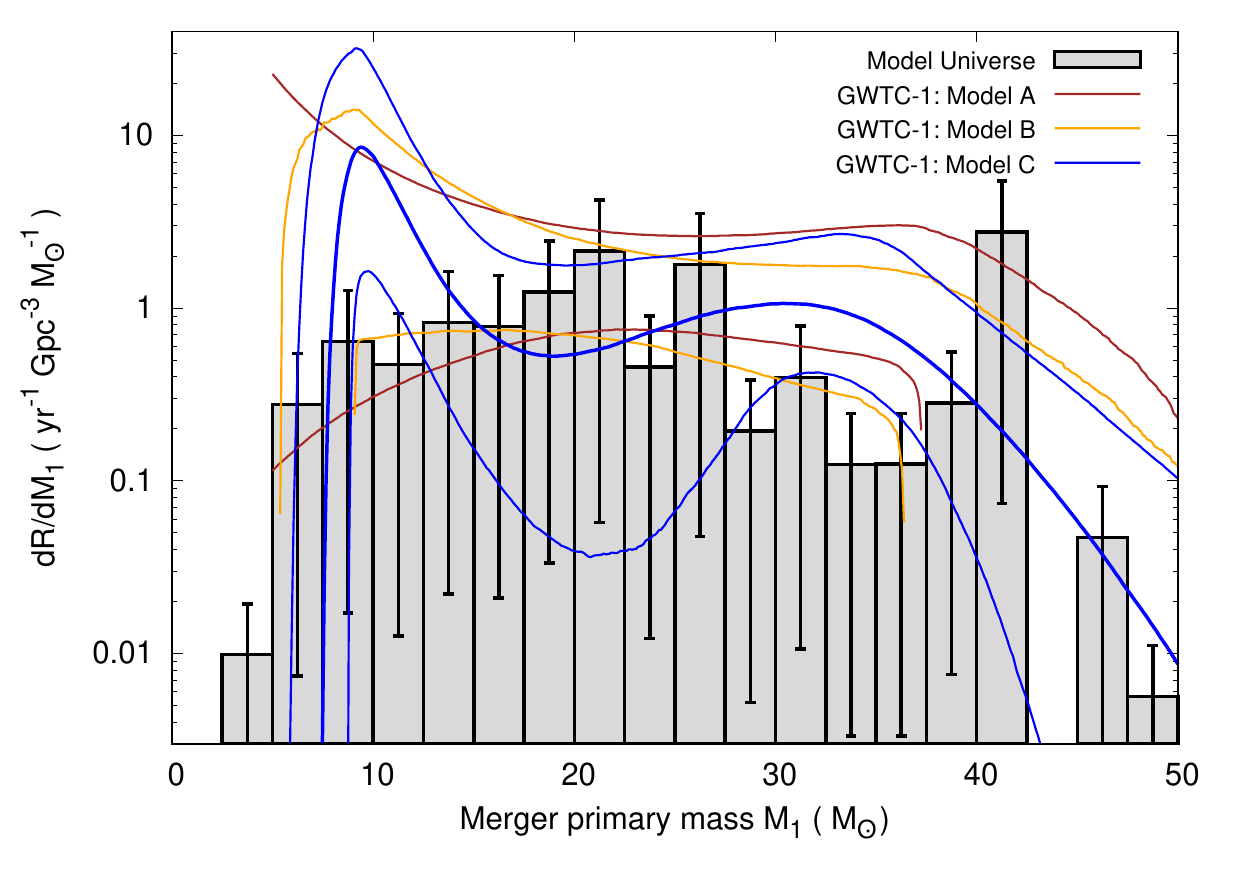}
\includegraphics[width=8.7cm,angle=0]{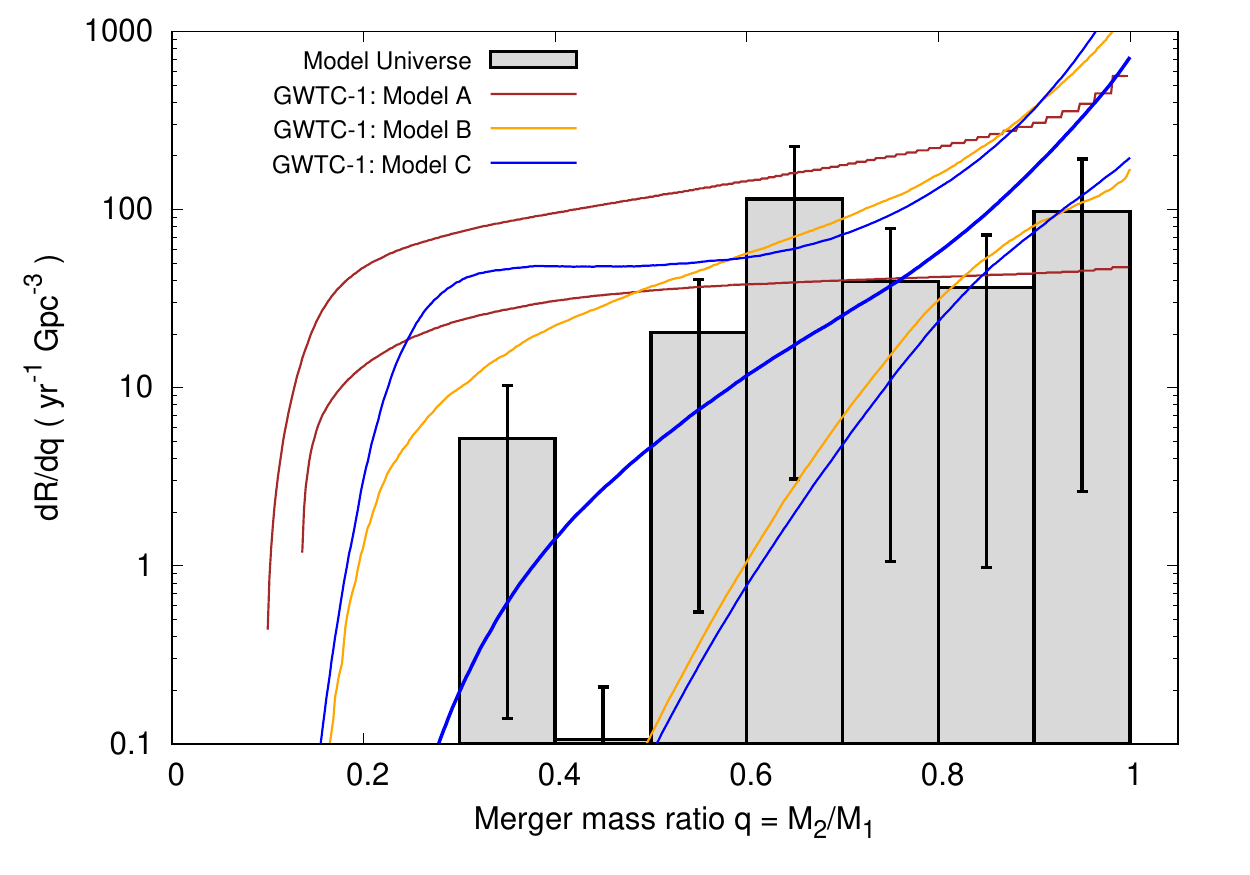}
\caption{{\bf Upper panels:} The filled histogram gives the present-day, differential intrinsic
merger rate density (Y-axis), as obtained from the Model Universe cluster population (Sec.~\ref{ratecalc}),
as a function of merger primary mass (left panel) and mass ratio (right panel) along the X-axis.
The upper and lower limits (histogram error bars) represent the reference and the
pessimistic rates (Sec.~\ref{ratecalc}, Table~\ref{tab_rates}), respectively. The heights of the histogram boxes
lie halfway between these two values (at approximately half of the reference value).
A visibility boundary (for average source inclination) at a redshift of $\zmax=1.0$ is assumed.
The solid lines are the differential intrinsic BBH merger rate densities as published
in the LVC GWTC-1 public repository, corresponding to their Model A, B, and C (legend) for BH masses
in merging BBHs \citep{Abbott_GWTC1_prop}. For each BH-mass model, the upper and lower
lines enclose the 90\% symmetric credible intervals; for Model C, the thick, blue line
gives the median.
{\bf Lower panels:} The same as in the top panels but for $\zmax=2.0$.
}
\label{fig:diffrate}
\end{figure*}

\begin{figure*}
\includegraphics[width=8.7cm,angle=0]{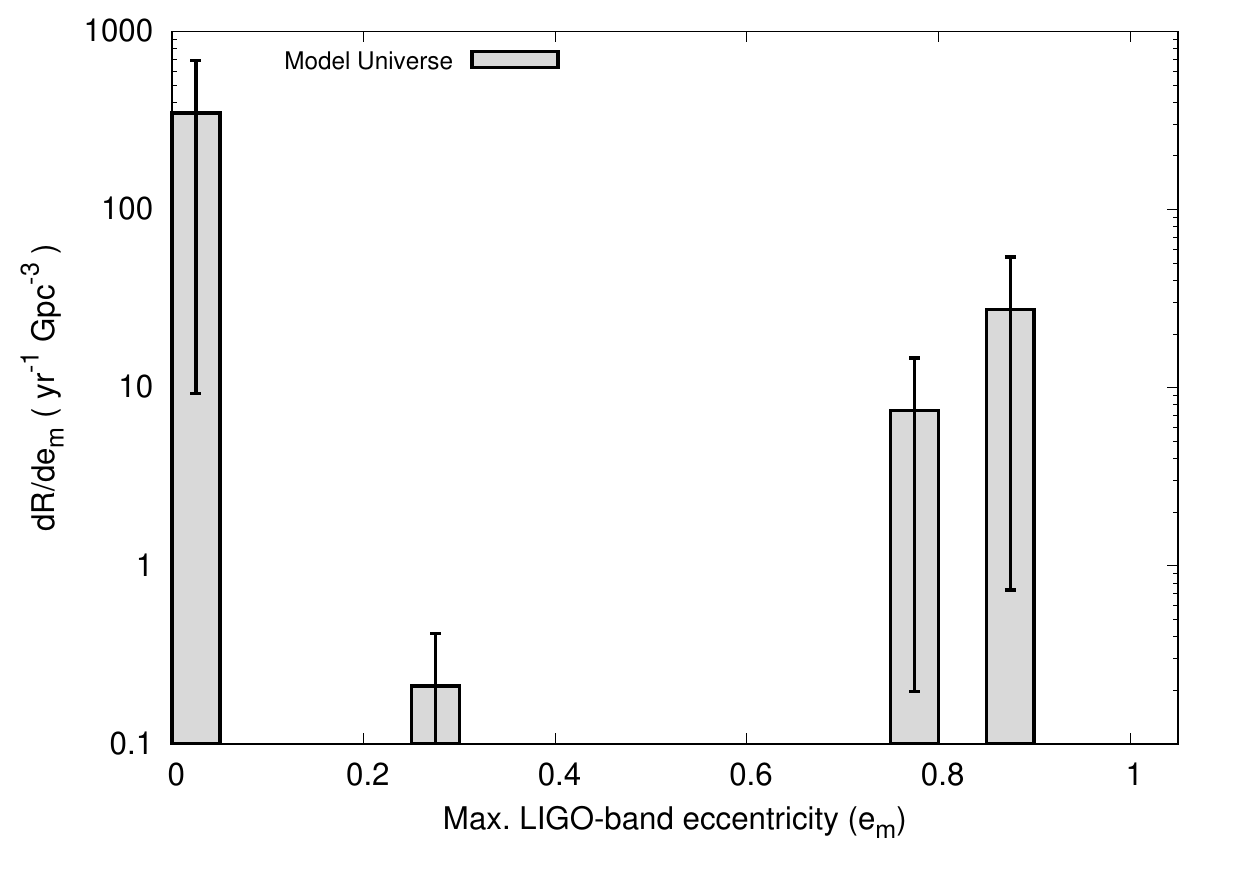}
\includegraphics[width=8.7cm,angle=0]{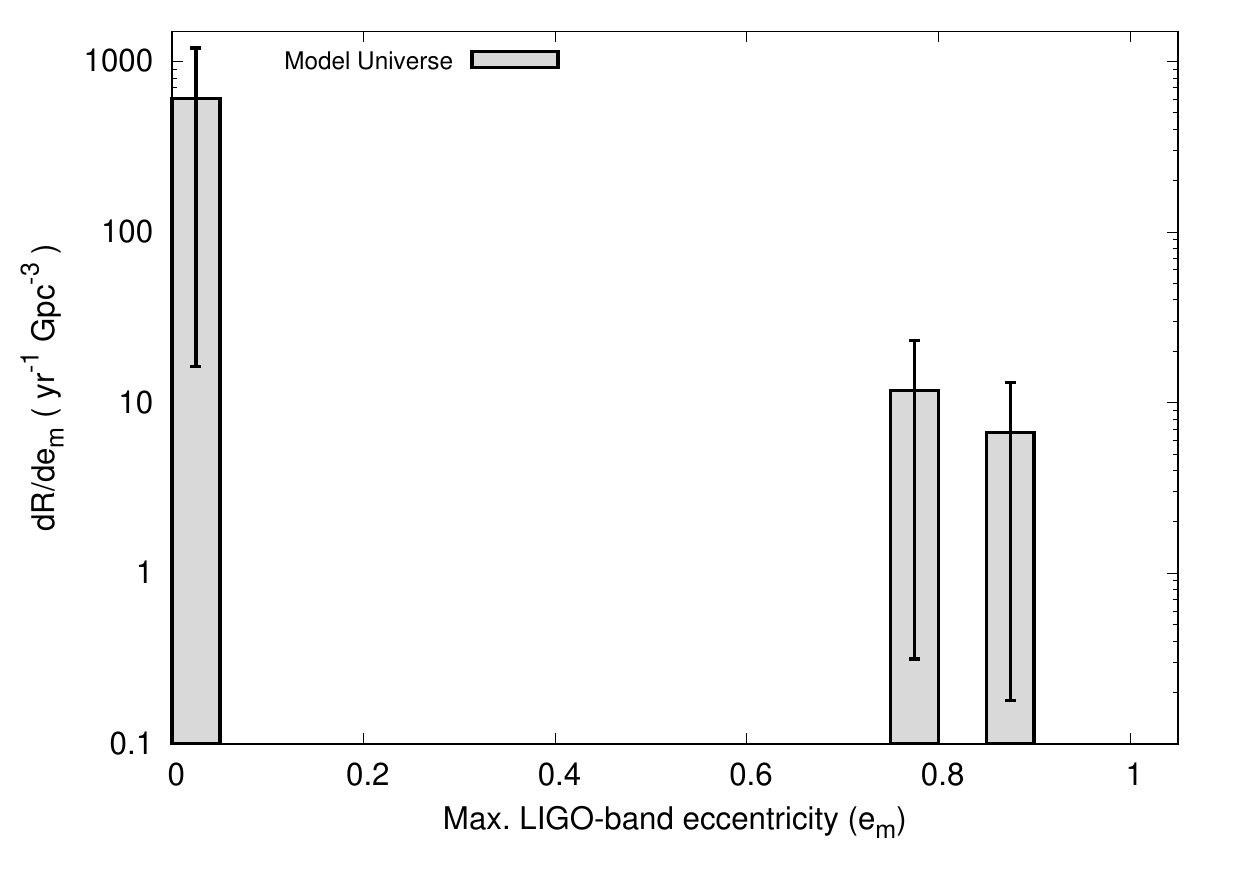}
\caption{The filled histograms give the differential
merger rate density (Y-axis), as obtained from the Model Universe cluster population (Sec.~\ref{ratecalc}),
as a function of maximum LIGO-Virgo-KAGRA-band eccentricity (Sec.~\ref{sec.diffr}), $\emx$, along the X-axis.
The upper and lower limits (histogram error bars) represent the reference and the
pessimistic rates (Sec.~\ref{ratecalc}, Table~\ref{tab_rates}), respectively. The heights of the histogram boxes
lie halfway between these two values (at approximately half of the reference value).
The left (right) panel corresponds to the visibility boundary (for average source inclination)
at the redshift $\zmax=1.0$ ($\zmax=2.0$).
}
\label{fig:drde}
\end{figure*}

\section{Results: merger rate density from young massive and open stellar clusters}\label{result}

Table~\ref{tab_rates} provides the present-day merger counts, $\nmrg$, out of
$\nsamp=5\times10^5$ Model-Universe clusters and the corresponding reference (pessimistic)
merger rate density, $\rate$ ($\rpess$), in mock detection experiments (Sec.~\ref{ratecalc})
with the detector horizon redshift, $\zmax$, varied from $1.0$ to $10.0$.

\subsection{Differential merger rate density}\label{sec.diffr}

The grey-filled histograms in the panels of Fig.~\ref{fig:diffrate} show the differential merger rate
densities, $d\rate/d\mone$ ($d\rate/dq$), w.r.t. the merger primary mass (mass ratio), $\mone$
($q\equiv\mtwo/\mone$; $\mone\geq\mtwo$), as obtained from the Model Universe.
The upper panel is the outcome with $\zmax=1.0$,
which visibility boundary is relevant for LVK O1, O2, and O3 observing runs \citep{Chen_2017b}.
The lower panel is for $\zmax=2.0$, relevant for future upgraded LIGO detectors such as (A+ and A++).
The thin-lined curves on the panels are the 90\% (symmetric) confidence limits on the BBH differential merger rate
densities as obtained, based on the GWTC-1 merger events, by \citet{Abbott_GWTC1_prop}
\footnote{The data for the differential merger rate densities are obtained from the 
public repository of GWTC-1 at \url{https://dcc.ligo.org/LIGO-P1800324/public}.}.
The limits corresponding to their Model A, B, and C, for BH masses in merging BBHs, are shown.
Among these, Model C is most consistent with the BH mass distribution in the present model clusters,
since, likewise Model C, the BH mass distribution is truncated and has a `bump' at $\approx40\Ms$
(due to PPSN; see, \eg, Fig.~8 of Paper I). Furthermore, likewise Model C, those BHs that
are retained in the clusters right after their birth (and, hence, can pair up dynamically) 
are of $\gtrsim10\Ms$ (see Fig.~8 of Paper I). Therefore, in this work, the comparisons
are done mainly with Model-C merger rate densities, the median value of which is also
shown in the panels of Fig.~\ref{fig:diffrate} (the thick, blue line). 

As seen in Fig.~\ref{fig:diffrate} (left panels), the reference $d\rate/d\mone$s from the Model Universe
(the histogram upper limits) well accommodate the median differential rate densities for Model C, for
most $\mone$ bins. The Model-Universe $d\rate/dq$s well accommodate the median Model-C differential
rate densities for $q\lesssim0.8$ (Fig.~\ref{fig:diffrate}; right panels).
The Model Universe tends to somewhat under-produce (near-)equal-mass mergers and over-produce unequal-mass
mergers, at the present day. This trend is a combination of the facts that
the dynamical channel in clusters is able to assemble unequal-mass merging BBHs, especially,
at shorter delay times, $\tmrg$, when the clusters contain a wider BH mass spectrum
(see Fig.~9 of Paper II), that
there are more mergers at shorter $\tmrg$ (see Fig.~\ref{fig:cosmorate}, right panel;
also Fig.~9 of Paper II),
and that the GW signals generated by the mergers at shorter $\tmrg$ ($\lesssim5.8$ Gyr) preferentially arrive the
detector at the present cosmic epoch, due to the adopted $\zmax\geq1.0$ detector horizon
and the consequent longer light travel times, $\tld$, from the clusters' comoving
distance, $D$
\footnote{Due to the assumed uniform spatial density of the clusters, the
probability density function of the clusters' distance redshift, $\zd$, increases
monotonically ($\propto D^2$, $D$ being the comoving distance corresponding to $\zd$)
as $\zd$ approaches the detector horizon, $\zmax$.}.

Overall, both Model-Universe $d\rate/d\mone$ and $d\rate/dq$ are consistent with the
90\% credible intervals of the GWTC-1 Model C differential merger rate densities (the thin, blue
lines in Fig.~\ref{fig:diffrate}), for both $\zmax=1$ and $2$.
The main exception to this is $d\rate/d\mone$ over $30\Ms\lesssim\mone\lesssim40\Ms$,
which falls below the Model-C lower limit. This is due to the inherent dearth of (direct-collapse) BHs, formed
(and retained right after birth) in the clusters, beyond $\approx30\Ms$ and before the `PPSN peak' at $40\Ms$
(see Fig.~8 of Paper I)\footnote{Such BH mass distribution occurs for metallicities $Z\lesssim\Zs/4$.
For $Z=\Zs$, the (retained) BH mass distribution truncates at $\approx15\Ms$ and no PPSN/PSN
takes place (see Paper I). However, for the cosmic metallicity evolution considered here \citep{Chruslinska_2019},
low $Z$ clusters form at all ages of the Universe.}.
This dearth is a result of the convolution of the ZAMS mass-remnant mass
relation (see, \eg, Fig.~6 of Paper I) with the standard stellar IMF adopted in these model clusters
(see Paper I for a detailed discussion). The same is true with $d\rate/d\mone$ and for the same reason,
when $\mone$ approaches $10\Ms$\footnote{The Model-Universe $d\rate/d\mone$ extends continuously below $\mone<5\Ms$,
\ie, close to, within, and below the (NS-BH) `mass gap',
due to the inclusion of the `delayed' remnant-mass scheme in some of
the models. A few of such models give rise to BBH mergers involving primaries close to and
within the mass gap. See Paper II for the details. $d\rate/d\mone$ below the mass gap occurs
due to the BNS mergers occurring in some of these models \citep{Fragione_2020b}.}.
As such, the mass distribution of BHs, retained in the model clusters right after their birth, is far 
from being (see Fig.~8 of Paper I) a simple power law plus a Gaussian peak at $40\Ms$,
as idealized in Model C.

The Model-Universe $d\rate/d\mone$ extends beyond $\mone>40\Ms$
and consistently with the Model-C boundaries (see Fig.~\ref{fig:diffrate}).
Such BH mass is the outcome of star-star mergers in binaries (see Fig.~12 of Paper I),
star-BH mergers in binaries via the formation of BH-Thorne Zytkow
object (see Paper II), or (first-generation) BBH mergers (see Paper II). 
The Model-Universe $d\rpess/d\mone$ and $d\rpess/dq$ (the histogram lower limits
in Fig.~\ref{fig:diffrate}) are highly conservative estimates (see Sec.~\ref{ratecalc})
and they fall below the lower boundaries of the Model-C differential merger rate densities.
The differential rate densities are generally higher and more biased towards
larger $\mone$ for $\zmax=2.0$ than for $\zmax=1.0$ (see Fig.~\ref{fig:diffrate}; left panels).
With larger $\zmax$, the detector is able to receive signals from shorter $\tmrg$ mergers
(see above) which mergers are generally more massive and numerous (see Figs.~4 \& 9 of Paper II),
causing such trends. The dependence of $\rate$ on $\zmax$ is further discussed
in Sec.~\ref{sec.rz}.

Fig.~\ref{fig:drde} plots the $d\rate/d\emx$ and $d\rpess/d\emx$ (the histogram upper and lower
limits, respectively) as obtained from the Model Universe cluster population. Here,
$\emx$ is the maximum LVK-band eccentricity, defined as the eccentricity
of the in-spiralling binary when its detector-frame (peak-power) GW frequency is 10 Hz or
the eccentricity at the minimum (peak-power) GW frequency, $f_{\rm min}$, if the binary's final inspiral
towards the merger begins at $>10$ Hz ($f_{\rm min}\sim10-100$ Hz in the detector frame,
for such LVK eccentric inspirals obtained in the present models;
see Fig.~10 and associated discussions in Paper II). In the panels of Fig.~\ref{fig:drde},
$\rate$ ($\rpess$) is nearly wholly concentrated over the bin at
the smallest $\emx$ ($0.00\leq\emx\leq0.05$).
Only a handful of binaries show up as eccentric (eccentricity $>0.1$) LVK mergers
at the present epoch, summing up to an eccentric-merger rate density of
$\recc\approx5.0\peryg$ ($\recc\approx2.5\peryg$) for $\zmax=1.0$; Fig.~\ref{fig:drde}, left panel
($\zmax=2.0$; Fig.~\ref{fig:drde}, right panel).
This $\recc$, being $\lesssim10$\% of $\rate$, is consistent with the rate of such eccentric mergers estimated
for GCs \citep[\eg,][]{Rodriguez_2018,Samsing_2018}. 

\begin{figure*}
\includegraphics[width=12.0cm,angle=0]{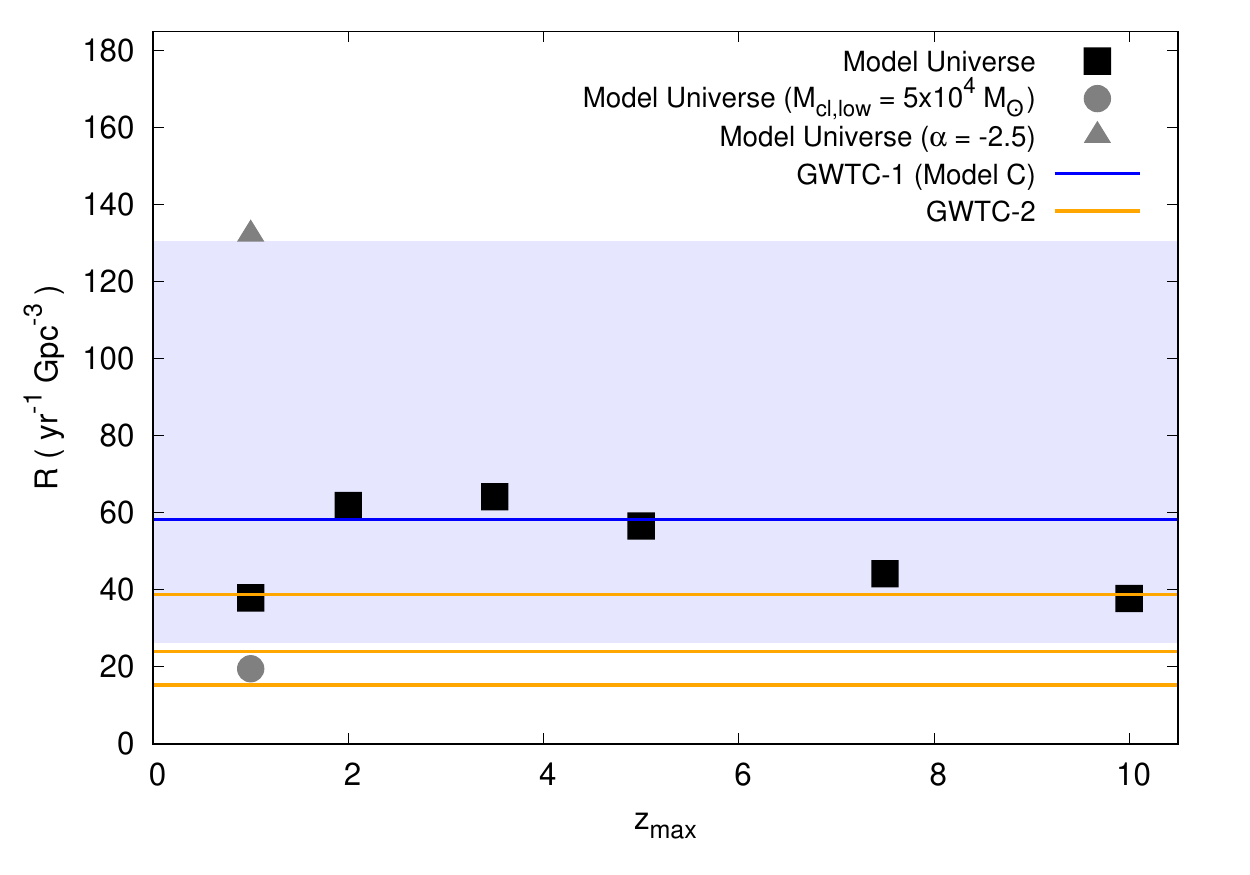}
\caption{The black, filled squares give the reference present-day intrinsic merger rate density (Y-axis),
from the Model Universe cluster population (Sec.~\ref{ratecalc}), as a function of the detector visibility
boundary (for average source inclination) redshift, $\zmax$ (X-axis). The grey, filled
triangle gives the merger rate density, for $\zmax=1$, for a Model Universe cluster population
with the cluster initial mass function being a power law of index $\alpha=-2.5$ (Sec.~\ref{discuss}). 
The grey, filled circle gives the merger rate density, for $\zmax=1$, for a Model Universe cluster population
with the cluster initial mass function truncated at the lower limit of
$\mcllow=5\times10^4\Ms$ (Sec.~\ref{discuss}). The
blue line indicates the median value of the BBH merger rate density estimated from GWTC-1 \citep{Abbott_GWTC1_prop},
for their Model C, and the blue-shaded background represents the corresponding 90\% credible interval.
The three orange lines are the median and the 90\% credible limits for the BBH merger rate density estimated
from GWTC-2 \citep{Abbott_GWTC2_prop}.}
\label{fig:rzmax}
\end{figure*}

\begin{figure*}
\includegraphics[width=8.7cm,angle=0]{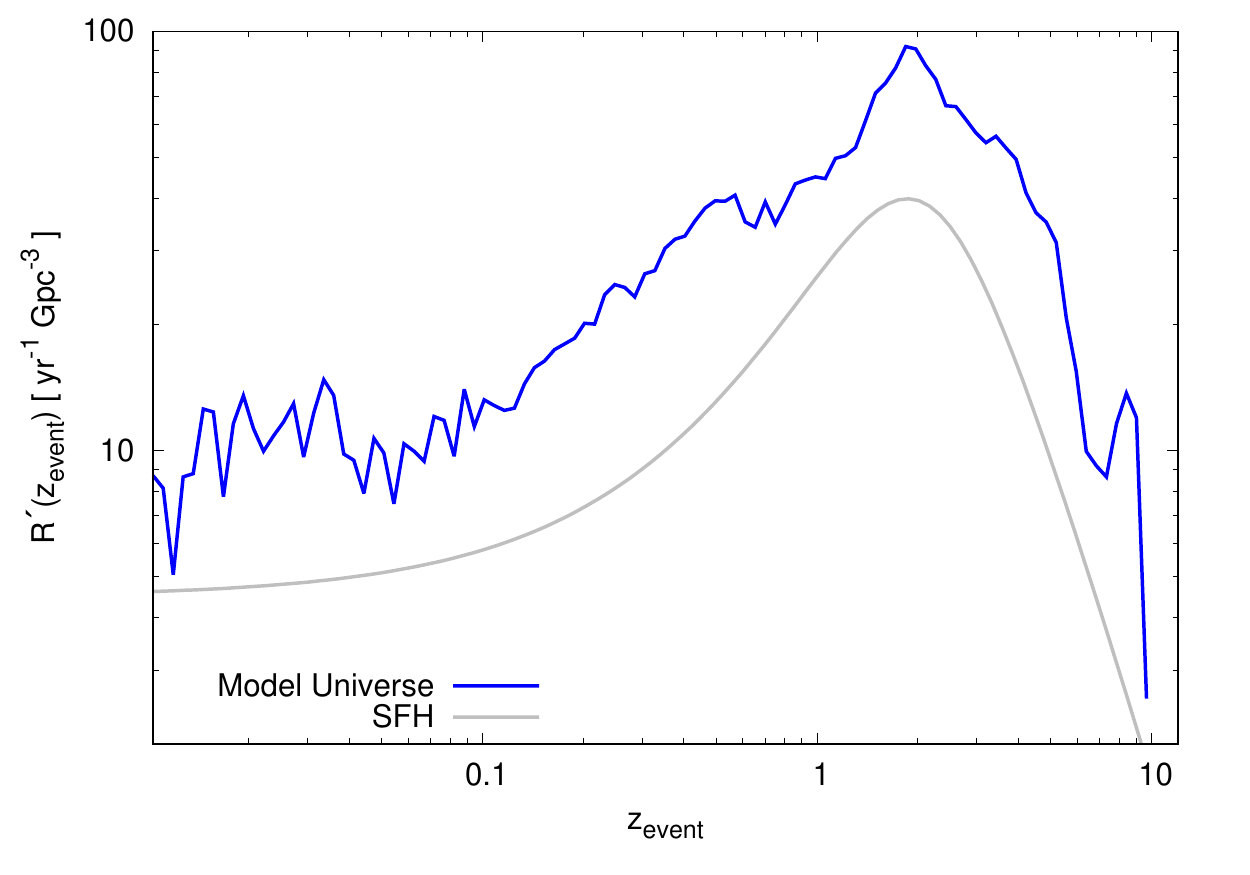}
\includegraphics[width=8.7cm,angle=0]{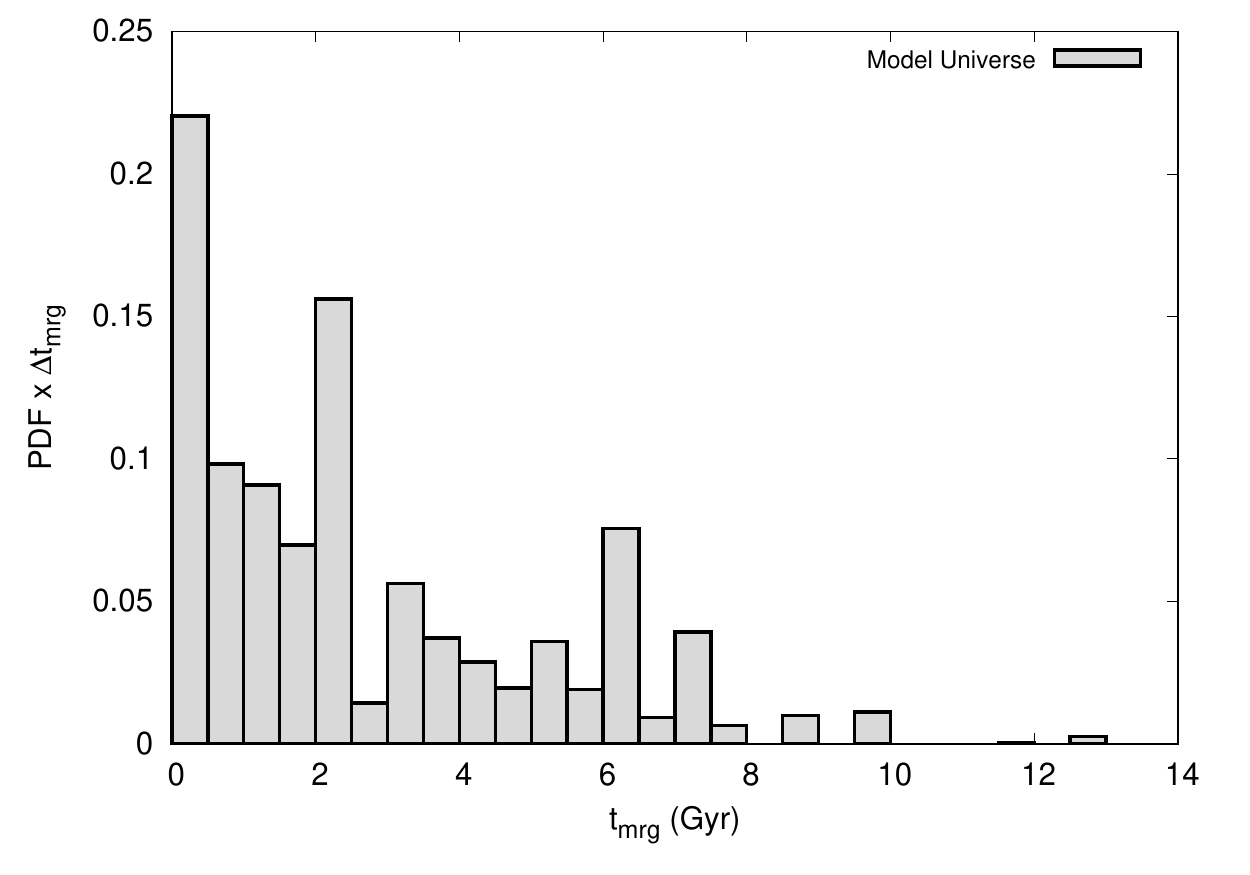}
\caption{{\bf Left panel:} The blue line gives the evolution of the (reference) cosmic merger rate density
($\rz$; Y-axis) with merger-event redshift, $\zevnt$ (X-axis), as obtained from the Model Universe
cluster population (Sec.~\ref{sec.rz}).
For visual comparison, the gray line shows the variation of cosmic star formation rate with redshift
\citep[][not to scale along the Y-axis]{Madau_2014}. {\bf Right panel:} the
delay time ($\tmrg$) distribution of the Model Universe cluster population. The $\rz(\zevnt)$
function and the $\tmrg$ distribution are constructed based on a sample of
$\nsamp=10^6$ clusters. To construct $\rz(\zevnt)$, $0\leq\zevnt\leq10$ is
divided into 100, equal-sized $\log_{10}$ bins.}
\label{fig:cosmorate}
\end{figure*}

\subsection{Redshift dependence of merger rate density}\label{sec.rz}

Fig.~\ref{fig:rzmax} plots (black, filled squares) the dependence of $\rate$ (reference merger rate density)
on $\zmax$ (detector visibility boundary) as obtained from the Model Universe;
these are the same entries as in Table~\ref{tab_rates}. Despite the Model Universe
undergoes the same star-formation and metallicity evolutions everywhere (as required
by the homogeneity and isotropy of the Universe), $\rate$ varies moderately
with $\zmax$, reaching a maximum at $\zmax\approx3.5$. As explained in Sec.~\ref{sec.diffr},
with increasing $\zmax$, one is able to reach to shorter merger delay times, $\tmrg$.
Since the $\tmrg$ distribution from the Model Universe cluster population
shows an overall increasing trend with decreasing $\tmrg$ (Fig.~\ref{fig:cosmorate}, right
panel), $\rate$ increases with increasing $\zmax$ (Fig.~\ref{fig:rzmax}; black, filled squares).
However, this competes with the fact that 
beyond $\zmax\approx1.85$, \ie, the redshift at which the cosmic SFH peaks
(Eqn.~\ref{eq:mdsfr}; Fig.~\ref{fig:cosmorate}, left panel, grey line),
the majority of the clusters form too late for
the long light travel times from comoving distances approaching $\zmax$.
Indeed, in Fig.~\ref{fig:rzmax} (see also Table~\ref{tab_rates}), $\rate$ increases only
slightly\footnote{The Poisson error in $\nmrg$ leads to merger-rate
uncertainties of $\Delta\rate\sim10^{-1}\peryg$ ($\Delta\rpess\sim10^{-3}\peryg$).}
as $\zmax$ increases from 2.0 to 3.5, beyond which $\rate$ begins to decline.    
In this way, the maxima of $\rate$ at $\zmax\approx3.5$ is an outcome of
two competing effects and is specific for the $\tmrg$ distribution obtained
from the Model Universe cluster population (given the $\Lambda$CDM Universe
with its currently-determined parameters and the cosmic SFH; see Sec.~\ref{ratecalc}). 

Fig.~\ref{fig:rzmax} shows that for $\zmax=1$, which
detector horizon is relevant for LVK O1, O2, and O3,
the Model Universe reference $\rate$($=37.9\peryg$; the black, filled square) falls moderately below
the median BBH merger rate density estimated from GWTC-1 \citep{Abbott_GWTC1_prop},
for their Model C ($=58.3\peryg$; blue line), but lies well within the corresponding 90\% credible
interval (blue-shaded area). The rather broad GWTC-1 limits accommodate the Model Universe reference
$\rate$s (black, filled squares) for all $\zmax$. On the other hand, the Model Universe $\rate$ at $\zmax=1$ 
nearly coincides with the 90\% credible upper limit of the significantly more
constrained GWTC-2 BBH merger rate density (\citealt{Abbott_GWTC2_prop}; orange lines).
This means that with suitable choices of astrophysical quantities in Eqn.~\ref{eq:rate},
\eg, $\mgclow$, $\mcllow$, $\rhogc$, and $\rmort$, the Model Universe cluster population
can reproduce the GWTC-2 median BBH merger rate density. This is discussed
further in Sec.~\ref{discuss}.

It would be worth looking into the inherent dependence of merger rate density
on merger-event redshift, $\zevnt$ (Sec.~\ref{ratecalc}),
as obtained from the Model Universe cluster population. This cosmic
merger rate density function, $\rz(\zevnt)$, is shown in Fig.~\ref{fig:cosmorate}
(left panel, blue line). Note that $\rz(\zevnt)$ is simply the collective
redshift distribution of the merger events from the Model Universe cluster population,
without taking light travel times into account, as opposed to
$\rate(\zmax)$ in Fig.~\ref{fig:rzmax} (see above). The $\rz(\zevnt)$ function
in Fig.~\ref{fig:cosmorate} is constructed based on a sample
cluster population (Sec.~\ref{ratecalc}) of $\nsamp=10^6$, the mergers
from which are distributed among 100, equal-sized $\log_{10}(\zevnt)$ bins, 
over $0\leq\zevnt\leq10$. The merger rate density at each bin is then
obtained from Eqn.~\ref{eq:rate} with $2\delobs$ replaced by $\delage$,
where $\delage$ is the Universe-age difference corresponding to
the redshift difference across the bin. The reference values of
$[\mgclow,\mgchigh]$ and $\rhogc$ are applied (Sec.~\ref{ratecalc}), \ie,
$\rz(\zevnt)$ in Fig.~\ref{fig:cosmorate} is the reference cosmic merger rate density.
The distribution of the merger delay times, $\tmrg$s, for the
Model Universe, as obtained from this sample cluster population,
is also shown in Fig.~\ref{fig:cosmorate} (right panel).
Due to the predominance of shorter $\tmrg$, $\rz(\zevnt)$ has a global peak at a
redshift $\zevnt=\zpeak$ very close to the cosmic SFH peak redshift,
\ie, at $\zpeak\approx1.85$ (Fig.~\ref{fig:cosmorate}, left panel)
\footnote{The somewhat spiked nature of $\rz(\zevnt)$, especially at low $\zevnt$,
is caused by the underlying relatively low, $\approx120$, total number of merger events
in the computed model clusters (see Table~C1 of Paper II). This also contributes to the
fluctuations in the $\tmrg$ distribution. The distinct peak of $\rz(\zevnt)$
at $\zevnt\approx10$ (Fig.~\ref{fig:cosmorate}, left panel) is caused by the distinct
peak in the smallest $\tmrg$ bin (Fig.~\ref{fig:cosmorate}, right panel).};
see also \citet{Santoliquido_2020}.
In contrast, $\rate(\zmax)$ peaks at $\zmax\approx3.5$ (see above; Fig.~\ref{fig:rzmax}).
Since within a volume enclosed by a spherical boundary at $\zmax$ the
sources are located closer to $\zmax$ with a higher probability (for a uniform
spatial density; see Sec.~\ref{sec.diffr}), $\rate(\zmax)$ will lie in between
$\rz(0)$ ($\rz(\zpeak)$) and $\rz(\zmax)$ for $\zmax\leq\zpeak$ (for $\zmax>\zpeak$).
Hence are the $\rate(\zmax)$ values (Fig.~\ref{fig:rzmax}).

\begin{figure*}
\includegraphics[width=8.7cm,angle=0]{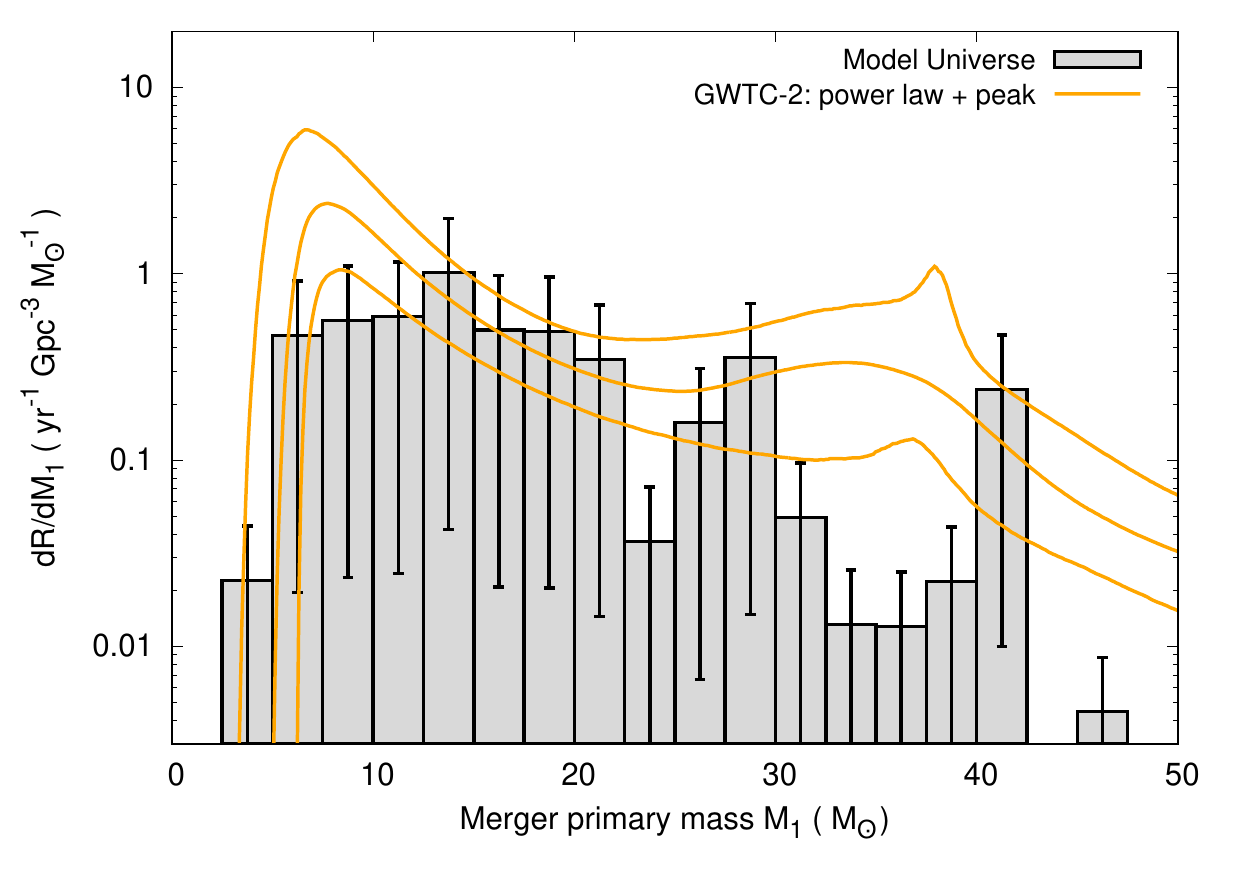}
\includegraphics[width=8.7cm,angle=0]{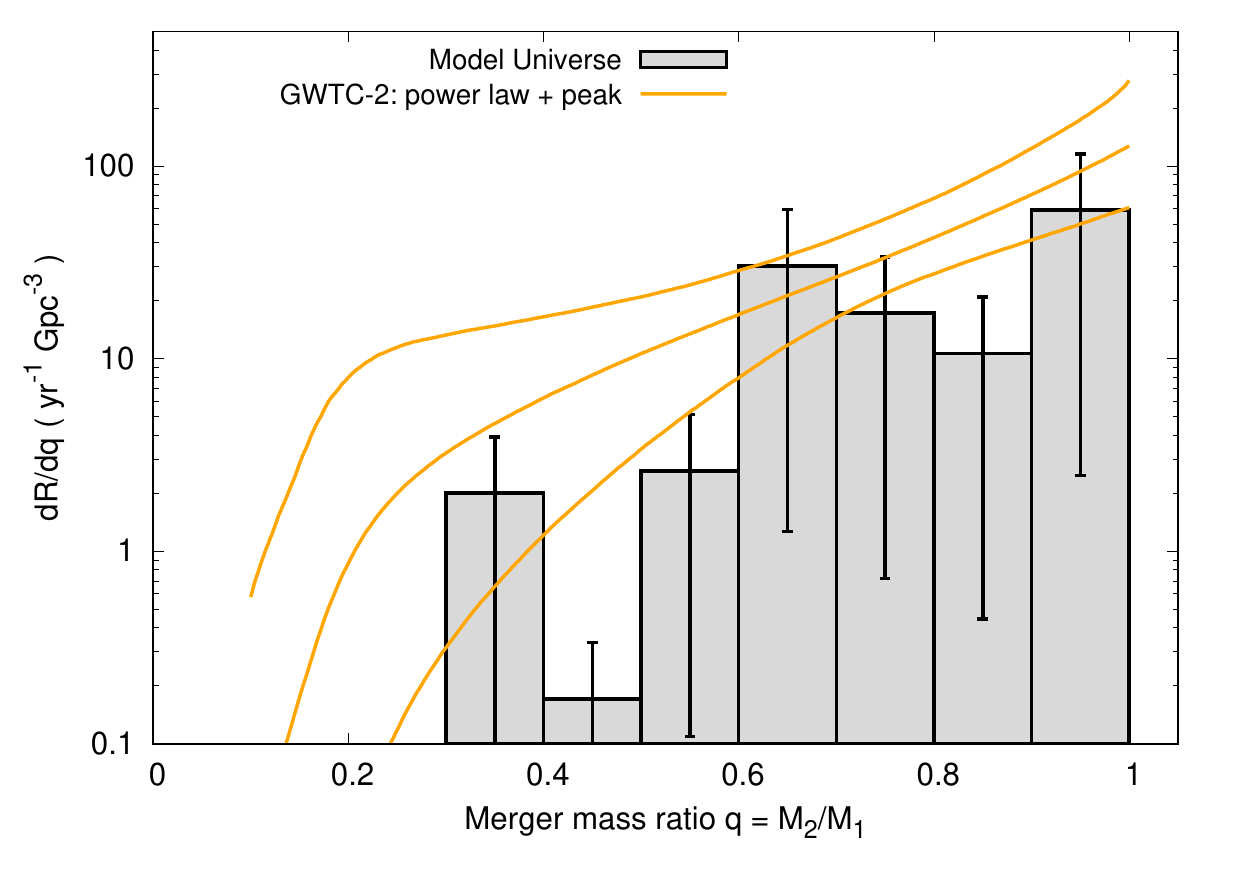}
\caption{The histograms are the same as in the top panels of Fig.~\ref{fig:diffrate} (\ie, corresponds to
the default sample cluster population with $\zmax=1$; see Sec.~\ref{ratecalc}, Table~\ref{tab_rates})
but renormalized so that the
reference values (histogram upper limits) sum up to the median BBH merger rate density as estimated
from GWTC-2 \citep{Abbott_GWTC2_prop}. The moderate renormalization corresponds to applying a somewhat smaller
GC-progenitor lower mass limit, $\mgclow=3.9\times10^5\Ms$ (Sec.~\ref{discuss}), in obtaining the reference $\rate$
(using Eqn.~\ref{eq:rate}) instead of the default reference lower limit ($\mgclow=5\times10^5\Ms$; Sec.~\ref{ratecalc}). 
The orange lines are
the median (central line) and the 90\% credible limits (upper and lower lines) of the differential BBH
merger rate density as obtained for GWTC-2 by \citet[][their `power law + peak' model]{Abbott_GWTC2_prop}.}
\label{fig:diffrate2}
\end{figure*}

\begin{figure*}
\includegraphics[width=10.0cm,angle=0]{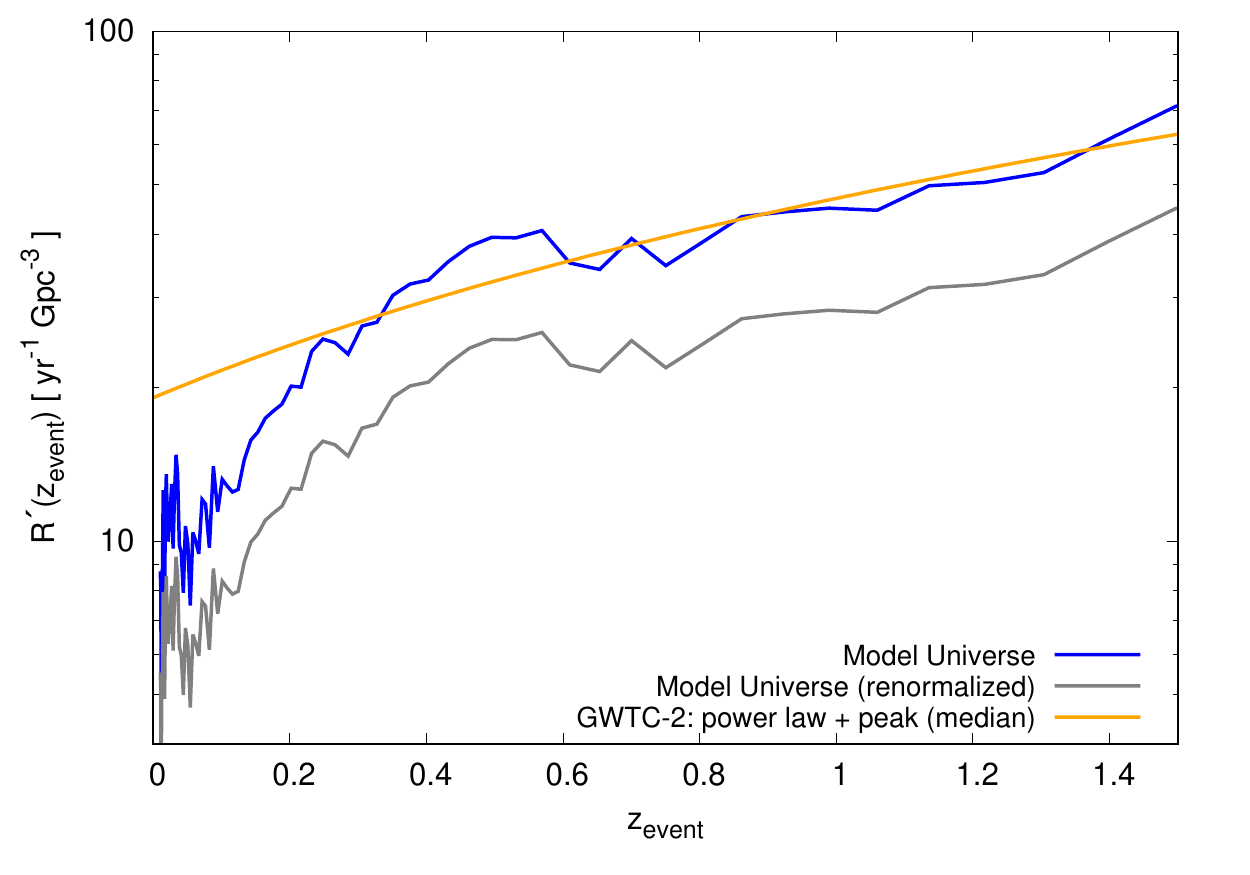}
\caption{The blue curve is the same as in Fig.~\ref{fig:cosmorate} but plotted up to
	$\zevnt=1.5$. The grey curve re-plots it with the renormalization described in
	Sec.~\ref{discuss} (see also the description in Fig.~\ref{fig:diffrate2}). The orange
	curve is the median cosmic merger rate density evolution as obtained from
	GWTC-2 by \citet[][their `power law + peak' model]{Abbott_GWTC2_prop},
	which curve is given by $\rz(\zevnt)=19.1(1+\zevnt)^{1.3}$.}
\label{fig:cosmorate2}
\end{figure*}

\section{Discussions: uncertainties in merger rate density}\label{discuss}

The uncertainties in the (differential) merger rate density, as obtained
in this study, is mainly driven by the
various astrophysical limits and factors in Eqn.~\ref{eq:rate}.
While the `laws' that enter Eqn.~\ref{eq:rate}, \ie,
$\clmf(\mcl)$ and $\sfh(\zf)$ (plus, implicitly, the $\Lambda$CDM Universe with
the \citealt{Planck_2018} parameters that partly determines $\nmrg/\nsamp$),
are based on observations (see Sec.~\ref{ratecalc} and references therein),
$\rate$ strongly depends on $\mcllow$ and $\mgclow$, $\clmf$ being a
power law of index $\alpha=-2$. Also, $\rate$ simply proportionates with $\rhogc$.
The strong dependence on these quantities is clear from the  
large difference in value between the reference $\rate$ and its pessimistic counterpart,
$\rpess$ (see Sec.~\ref{ratecalc}; Table~\ref{tab_rates}).

Although $\alpha=-2$ is the `widely accepted' value of the cluster
birth mass function index (based mainly on photometric mass estimates of
young, gas-free clusters; see, \eg, \citealt{Gieles_2006b,Larsen_2009,PortegiesZwart_2010,Bastian_2012}),
observations also suggest potential moderate variations of $\alpha$ \citep[\eg,][]{Ryon_2015,Webb_2021}.
To probe the dependence of $\rate$ on moderate alterations of $\alpha$,
a sample cluster population is constructed as described in Sec.~\ref{ratecalc}
but out of a $\clmf$ with $\alpha=-2.5$ ($\zmax=1$ is assumed).
Despite the resulting $\nmrg/\nsamp$ is somewhat smaller compared
to that with $\alpha=-2$ (as expected,
since less massive clusters, which are more predominant for $\alpha=-2.5$,
tend to produce less number of mergers per cluster; see Table~C1 of Paper II),
the corresponding $\rate$ is $\approx3.5$ times higher
(compare between the $\zmax=1$ entries in Table~\ref{tab_rates}).
This rate is also indicated in Fig.~\ref{fig:rzmax} (the grey, filled triangle),
which exceeds the 90\% credible upper limit from GWTC-1.

It would also be of interest to examine the impact of the lower mass cutoff, $\mcllow$, of $\clmf$  
on $\rate$. With $\rmort=1$ (Sec.~\ref{ratecalc}, Eqn.~\ref{eq:rate}), $\mcllow$
serves as an effective cutoff: clusters either fail to assemble efficiently as gas-free,
gravitationally-bound young clusters or preferentially get
destroyed after successful assembly due to environmental effects (\eg,
interactions with molecular clouds) with initial masses below $\mcllow$.  
A sample cluster population is constructed with $\mcllow=5\times10^4\Ms$
but the other ingredients being as default (see Sec.~\ref{ratecalc})
and $\zmax=1$. Despite the resulting $\nmrg/\nsamp$ is nearly doubled
(as expected, since more massive clusters tend to produce a larger number
of mergers per cluster; see Table~C1 of Paper II),
the corresponding $\rate$ is nearly halved
(compare between the $\zmax=1$ entries in Table~\ref{tab_rates}).
This rate is indicated in Fig.~\ref{fig:rzmax} (the grey, filled circle),
which lies close to the GWTC-2 median value.

With the default sample cluster population (Sec.~\ref{ratecalc}),
$\rate$, at $\zmax=1$, is close to the 90\% credible upper limit
of the GWTC-2 BBH merger rate density (Sec.~\ref{sec.rz}, Fig.~\ref{fig:rzmax}). Nevertheless,
the median GWTC-2 value is obtained by calculating the reference $\rate$ 
from Eqn.~\ref{eq:rate} with a slightly lower $\mgclow=3.9\times10^5\Ms$
\footnote{Interestingly, this GC-progenitor lower limit is very close
to that in \citet{Antonini_2020b}.}
instead of the default reference lower mass limit of
GC progenitors ($\mgclow=5\times10^5\Ms$; Sec.~\ref{ratecalc}).
This altered $\mgclow$ is still consistent with being progenitors
of present-day GCs \citep{Kremer_2020}. Fig.~\ref{fig:diffrate2}
shows the Model Universe $d\rate/d\mone$ and $d\rate/dq$, for $\zmax=1$,
corresponding to the integrated $\rate$ matching the GWTC-2 median
rate ($23.9\peryg$). As seen in Fig.~\ref{fig:diffrate2}, the
Model Universe (renormalized) reference differential merger rate
densities agree reasonably with the GWTC-2 median and 90\%
credible limit merger rate densities, for their `power law + peak' BH mass model \citep{Abbott_GWTC2_prop}
\footnote{The data for the differential merger rate densities are obtained from the 
public repository of GWTC-2 at \url{https://dcc.ligo.org/LIGO-P2000434/public}.
A modified version of the {\tt Python} script provided in the directory
{\tt Fig-3-m1-ppd} is utilized to extract the relevant data from the dataset located
in the directory {\tt Multiple-Fig-Data}.}.
This prior for BH masses in merging BBHs is similar to the Model C prior (Sec.~\ref{ratecalc})
used to obtain the GWTC-1 differential merger rate densities \citep{Abbott_GWTC1_prop}.

The Model Universe $d\rate/d\mone$, for $\zmax=1$, is somewhat focussed
towards $\mone\gtrsim20\Ms$ whereas the corresponding GWTC-2 differential merger rate density
has a dominant peak below $\mone\lesssim10\Ms$ (Fig.~\ref{fig:diffrate2}, left panel).  
As discussed in Sec.~\ref{sec.diffr}, the mass distribution of BHs, which retain
in the present model clusters after birth and which pair-up via dynamical
interactions, is far from being a well defined, simple function such as
a power-law plus peak. With the inclusion of BBH merger events involving primaries
closer to the mass gap and a larger number of BNS mergers, the GWTC-2 differential
merger rate density extends down to $\mone$ that is very similar to that
from the Model Universe (see Fig.~\ref{fig:diffrate2}, left panel;
Sec.~\ref{sec.diffr}). It is clear that with increasingly improved
constraints on (differential) compact-binary merger rate density from LVK
observations, it would be possible to provide constraints on
widely debated issues regarding star cluster formation and large scale structure formation, \eg, 
mass dependence of cluster formation efficiency, $\rmort(\mcl)$, and
lower and upper mass limits, $[\mgclow,\mgchigh]$, of GC progenitors
\citep[see, \eg,][]{Rodriguez_2015,Banerjee_2018b,Kruijssen_2019,ElBadry_2019b,Krumholz_2019}.
The differential
merger rate density profiles would also help constraining the relative
contributions of the various other channels for producing compact
binary mergers, \eg, dynamical evolution of field hierarchical systems, mergers
via Kozai-Lidov mechanism in galactic nuclei, and evolution of
field massive binaries.

Fig.~\ref{fig:cosmorate2} compares the Model Universe reference cosmic merger rate density
evolution, $\rz(\zevnt)$, with the median cosmic merger rate density evolution from GWTC-2 \citep{Abbott_GWTC2_prop}.
With both normalizations (\ie, that in Sec.~\ref{ratecalc} and that in this section), the Model
Universe curves lie close to the GWTC-2 curve for $\zevnt\gtrsim0.2$ (with the scaling
in this section, the Model Universe curve is $\approx40\%$ lowered along the Y-axis). Both of
the Model Universe $\rz(\zevnt)$ lie well within the rather broad 90\% credible
intervals of the GWTC-2 cosmic merger rate density \citep[see Fig.~14 of][]{Abbott_GWTC2_prop}.

In this context, it is important to note that the present (differential) merger rate density estimates,
based on the Model Universe sample cluster population (Sec.~\ref{ratecalc}),
do not assume any absolute (cosmic) star cluster formation efficiency as a result of the
cosmic star formation process(es). What goes
into the rate estimates is the number of gas-free, bound clusters \emph{forming} per unit comoving volume,
beyond a certain threshold (birth) mass $\mcllow$, \emph{relative} to the observed number
density of present-day GCs, $\rhogc$. In other words, the formation
efficiency of lower mass clusters relative to the massive progenitor clusters of what we  
now call GCs. The present treatment also does not require all clusters
to survive until the present cosmic epoch \citep[see also][]{Rodriguez_2018b,Fragione_2018b,Antonini_2020b}.
An inherent assumption, however, is that the present-day GCs are the relics of the same
cosmic star and cluster formation laws that continue to shape the present-day
cluster mass function (the same $\sfh$ goes into the numerator and denominator
of Eqn.~\ref{eq:rate}). However, GCs, being formed during the galaxy assembly
phase, may have undergone a very different SFH.
The Model Universe merger rate density is proportional to
the spatial density of GCs, $\rhogc$ (Eqn.~\ref{eq:rate}). Interestingly,
the $\rhogc=2.57\permv$ used for obtaining the reference $\rate$
\citep[][see Sec.~\ref{ratecalc}]{PortegiesZwart_2000} is very close to
what can be expected from dark matter halo mass density from numerical
simulations of large scale structure formation and the observed relation between
the total GC-population mass and the dark matter halo mass in galaxies
\citep{Antonini_2020b}.

Boost factors similar to those applied in Eqn.~\ref{eq:rate} (Sec.~\ref{ratecalc}) 
are also incorporated in recent studies, \eg, \citet{Rodriguez_2018b,Fragione_2018b,Antonini_2020b}. 
The work of \citet{Antonini_2020b} suggests a boost factor of $32.5^{+29.8}_{-12.5}$
in mapping the present-day GC mass density
to its `birth' or `initial' value (based on a Markov Chain Monte Carlo approach).
This large boost incorporates the conversion of
an $\alpha=-2$ power law (or Schechter; \citealt{Gieles_2006b,Larsen_2009}) initial GC mass function
to the present-day, observed peaked GC mass distribution \citep{Harris_1996}, through various
mass loss and destruction mechanisms (stellar-evolutionary mass loss, evaporation of stars
driven by two-body relaxation, tidal stripping). For evaluating the reference $\rate$ here, the boost
factor due to the $\clmf$ integrals (Eqn.~\ref{eq:rate}; Sec.~\ref{ratecalc}) is
$\approx40.0$, being consistent with that in \citet{Antonini_2020b}
\footnote{This is expected since $[\mcllow,\mclhigh]$ for the sample
cluster population lies below the `turn over' mass of the present-day GC mass distribution,
which mass range primarily gets depleted in the `turning over' process
and, hence, dominantly contributes to the boost factor.}
(for $\rpess$, the factor is $\approx4.2$). The present estimate incorporates
an additional boost factor of $\approx3.4$ due to YMCs and OCs undergoing a
much longer SFH, up to $z=0$, in contrast to the progenitors of present-day GCs
(Eqn.~\ref{eq:rate}; Sec.~\ref{ratecalc}). The time evolution of the
cluster mass function, beginning from $\clmf$, is naturally incorporated in the
present work since the sample cluster population is built based on long term
evolutionary model clusters (Sec.~\ref{ratecalc}; Paper II).

\begin{figure*}
\includegraphics[width=8.7cm,angle=0]{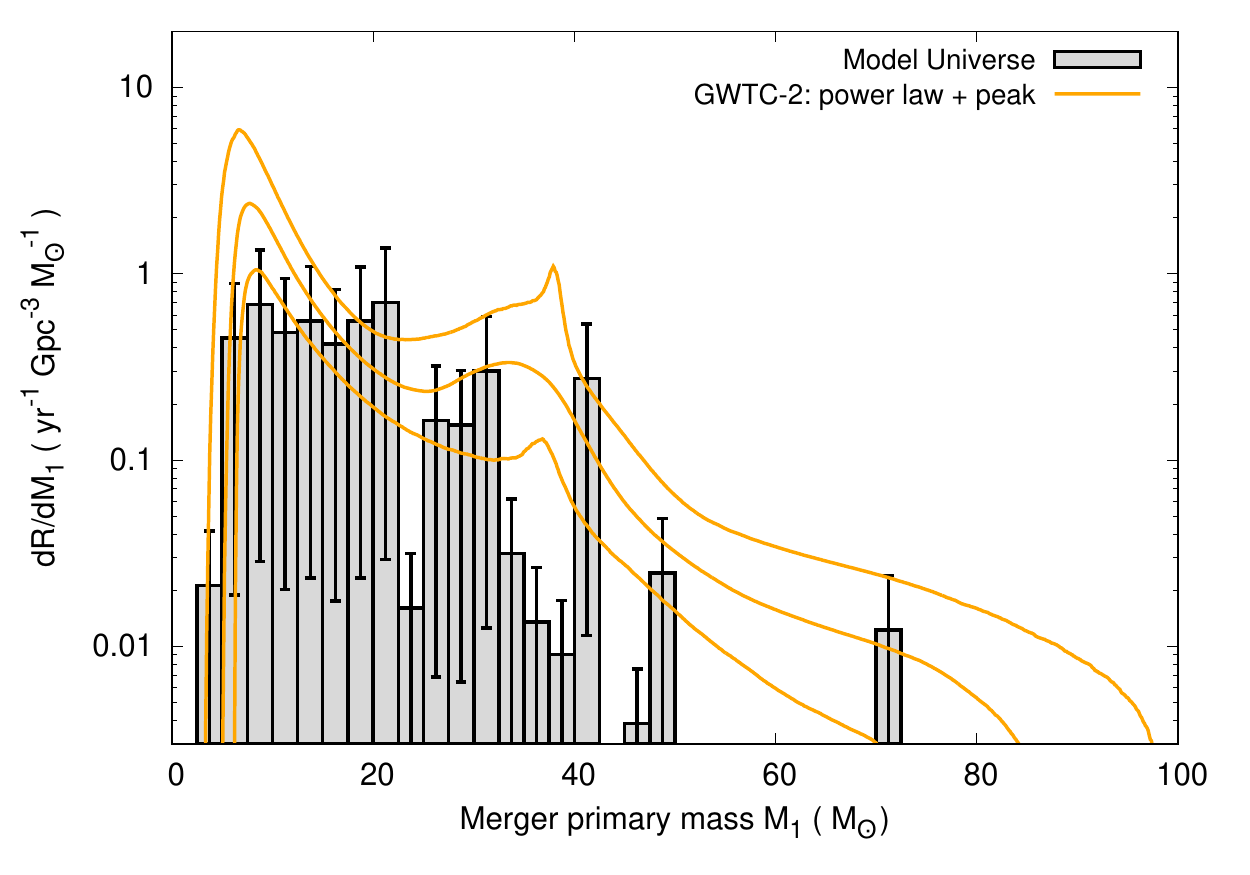}
\includegraphics[width=8.7cm,angle=0]{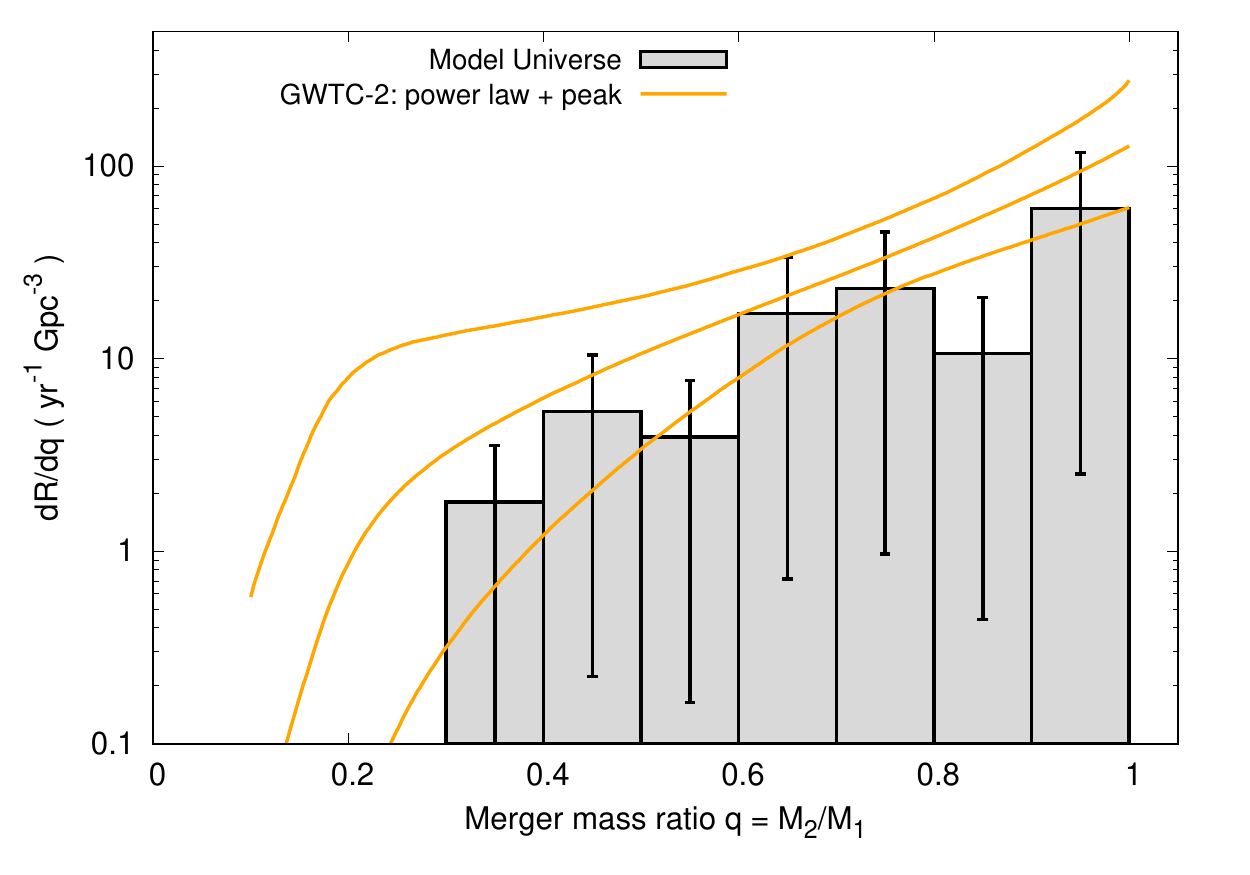}
	\caption{The histograms are the same as in the panels of Fig.~\ref{fig:diffrate2}, with the same renormalization
	applied, but the sample cluster population is constructed
	utilizing the evolutionary cluster model set shown in Table~\ref{tab_nbrun} (instead of those in Table C1
	of Paper II).}
\label{fig:diffrate3}
\end{figure*}

The Model Universe merger rate densities and differential merger rate density profiles, as obtained
from the present computed cluster model sets, has potentially been affected
by the incompleteness and inhomogeneities in the model grid (Paper II).
Nevertheless, it is ensured that the intended $\clmf(\mcl)$
and $\sfh(\zf)$ are reasonably achieved for each sample cluster population.
To demonstrate the effect of inhomogeneity and incompleteness of
the model grid with which the sample cluster population is constructed, an
independent mock detection experiment (default sample cluster population with $\zmax=1$)
is performed solely with the cluster evolutionary models
listed in Table~\ref{tab_nbrun}. This model set is richer in $\mcl=10^5\Ms$ models
(6 of them) which span over $0.001\leq Z\leq0.02$ and two of them having primordial
binaries. The lower $\mcl$ models in Table~\ref{tab_nbrun} are those from
Table~C1 of Paper II with $\mcl\geq2\times10^4\Ms$ and $\rh\leq2$ pc plus
two additional low-$Z$ $\mcl=2\times10^4\Ms$ models. In contrast to
Table~\ref{tab_nbrun}, Table~C1 of Paper II (with which model set all the previous experiments
in this paper are done) contains only one completed model with $\mcl=10^5\Ms$
and only the models with $\mcl=3\times10^4\Ms$ have $\rh=3$ pc (which $\rh=3$ pc models are excluded
in Table~\ref{tab_nbrun}). The outcome of the experiment with the Table~\ref{tab_nbrun}-set
is shown in the final row of Table~\ref{tab_rates}. The resulting merger rate
densities are only slightly, $\approx10\%$, smaller than those obtained with the set in Paper II.

Fig.~\ref{fig:diffrate3} shows the resulting Model Universe differential merger rate densities
with the same renormalization as in Fig.~\ref{fig:diffrate2}, \ie, the total Model Universe
reference merger rate density having been equated to the GWTC-2 median BBH merger rate density. Likewise in
Fig.~\ref{fig:diffrate2}, the Model Universe reference differential
merger rate densities in Fig.~\ref{fig:diffrate3} agree reasonably with the GWTC-2 median
and 90\% credible limit merger rate densities, for their `power law + peak' BH mass model, even
well within the PSN mass gap ($\mone>40\Ms$). The overall pattern of the Model Universe differential
merger rate densities using the Table~\ref{tab_nbrun} model set is similar to that
using the Paper II set, except that the former $d\rate/d\mone$ extends more
prominently within the PSN mass gap.

In the mock detection experiments presented here, a model cluster is chosen
irrespective of its primordial binary content. Due to the much higher computational
costs of N-body models with binaries, only about half of the models
(in both the Paper II and Table~\ref{tab_nbrun} sets)
contain primordial binaries (Sec.~\ref{nbcomp}). This makes the binary fraction
among O-type stars in the Model Universe to be $\approx50$\% which is still
consistent with observations \citep{Sana_2013} (the Model-Universe binary fraction among all
stars is $<5$\%). As argued in \citet[][see their Sec.~II. A.]{Banerjee_2020d},
the presence or absence of primordial binaries in a model cluster do not grossly influence
its dynamical GR-merger outcome, since a dynamically-active BH population inside a cluster
interacts inefficiently with the cluster's stars and stellar binaries. Hence, the present inhomogeneity in
the model sets regarding primordial binary content is unlikely to have significantly influenced the
BBH merger rates obtained here.

Nevertheless, a high binary fraction among O-type stars in a massive stellar cluster
can potentially influence the mass distribution of BHs, that retain in the cluster and participate in
dynamical mergers, due to dynamically-facilitated star-star mergers occurring in the clusters.
This effect becomes particularly prominent at low metallicities by causing a high mass
tail of the retained BHs' mass distribution that may extend well into the PSN gap
\citep[Paper I,][]{Spera_2019,Gonzalez_2020}. In fact, the introduction of the two
$Z=0.001$, $\mcl=10^5\Ms$ evolutionary models with primordial binaries in the set of Table~\ref{tab_nbrun}
is what primarily causes the Model Universe $d\rate/d\mone$ in Fig.~\ref{fig:diffrate3}
to extend into the PSN gap (as opposed to that in Fig.~\ref{fig:diffrate2} where these two models
are not used). This clearly demonstrates the importance of homogeneity and ergodicity of the
various physical parameters in the model cluster grid, in estimating the (differential)
merger rate density of the Universe. The gaps and inhomogeneities in the model
grid of Paper II are caused by the tediousness of N-body calculations, especially,
towards the high-mass end and with the inclusion of primordial binaries. It is
also due to the exploration of various alternative prescriptions in Paper II
(none of which would grossly affect the cluster evolution or its merger
outcome; see \citealt{Banerjee_2020d}).

The present model clusters do not incorporate any `relic' of the pre-assembly violent-relaxation
phase, \eg, initial substructures and initial mass segregation. However, such details
are unlikely to have an impact on the merger yield of the model clusters
as discussed in \citet[][Sec.~II. A. and references therein]{Banerjee_2020d}.

The present evolutionary model sets comprise one N-body computation per
model cluster and hence lack information about stochastic variations of the number
of mergers and of their delay times, for a given model. Hence, despite the small
Poisson errors ($\Delta\rate\sim10^{-1}\peryg$, $\Delta\rpess\sim10^{-3}\peryg$),
error propagation is not properly taken
into account in the present calculations to provide limits of $\rate$ and $\rpess$,
which will be taken up in a future study.

The present mock detection experiments do not incorporate detector sensitivity curves
and GW strain for the mergers' (luminosity) distances (Paper II), since only intrinsic
merger rate densities are evaluated. Such a mock detection experiment can be
straightforwardly extended to include a signal-to-noise-ratio threshold,
as described in \citet{Banerjee_2020d}, which will be taken up in a future study
to estimate the merger detection counts with ground-based GW detectors.

Although the $\Lambda$CDM cosmological framework is adopted in the present work,
any alternative Universe framework can also be incorporated in the current approach
with alternative interrelations between redshift, Universe age, and light travel time. 

\section{Summary and outlook}\label{summary}

This work estimates the present-day intrinsic merger rate density and
its derivatives (\ie, the differential merger rate densities) w.r.t. the merging
binary's primary mass, mass ratio, and eccentricity from GR compact binary mergers
occurring due to dynamical interactions in YMCs and OCs.
To that end, a set of computed model clusters, with up to date stellar-evolutionary and
stellar remnant formation schemes
and PN treatment of compact binary merger events (Paper I; Paper II; Sec.~\ref{nbcomp}), 
is utilized to construct sample cluster populations in a $\Lambda$CDM Model Universe (Sec.~\ref{ratecalc}).
From such a sample cluster population, merger GW signals are accumulated
at the present cosmic epoch ($z=0$) in an idealized LVK-type
detector (Sec.~\ref{ratecalc}). The model clusters, initially, have
masses spanning over $2\times10^4\Ms\leq\mcl\leq10^5\Ms$, half-mass radii over
$1{\rm~pc}\leq\rh\leq3{\rm~pc}$, and metallicity over $0.0001\leq Z \leq0.02$
and are composed of stars following a standard IMF with the O-type stars
being in an observationally-motivated distribution of primordial binaries
(Sec.~\ref{nbcomp}; Paper II; \citealt{Sana_2011,Moe_2017}). Such initial
model cluster properties are consistent with those observed in fully-grown,
(near-)gas-free, (near-)spherical YMCs. Each sample cluster
population is constructed (Sec.~\ref{ratecalc}) following initial cluster
mass distribution $[\clmf(\mcl)\sim\mcl^{-2}]$, cosmic SFH \citep{Madau_2014},
and cosmic metallicity evolution \citep{Chruslinska_2019} that are all
derived from observations. The member clusters of a sample (of size $\nsamp=5\times10^5$) 
are distributed in the Model Universe with a uniform spatial distribution
and within a detector visibility horizon located at a redshift $\zmax$.
That way, sample cluster populations are obtained with $\zmax$ varying
from 1.0 to 10.0, which cluster populations produce merger event counts,
$\nmrg$, at the present epoch (Sec.~\ref{result}, Table~\ref{tab_rates}).
The resulting $\nmrg/\nsamp$ values are then scaled to estimate the
merger rate densities and differential merger rate densities
of the Model Universe (Sec.~\ref{ratecalc}).

For $\zmax=1$, which represents the detector horizon for
LVK O1-O3 observing runs, the Model Universe reference (pessimistic)
BBH merger rate density is evaluated to be $\rate=37.9\peryg$ ($\rpess=0.51\peryg$) 
(Sec.~\ref{result}, Table~\ref{tab_rates}). This merger rate density range
well accommodates the GWTC-1 BBH merger rate density \citep{Abbott_GWTC1_prop}
and as well the much more constrained GWTC-2 BBH merger rate density \citep{Abbott_GWTC2_prop}, 
given the median values and the 90\% credible intervals of the GWTC estimates
(Sec.~\ref{sec.rz}; Fig.~\ref{fig:rzmax}). Also, the $\zmax=1$ Model Universe differential
merger rate densities w.r.t. merger primary mass and mass ratio
($[d\rpess/d\mone,d\rate/d\mone]$ and $[d\rpess/dq,d\rate/dq]$, respectively)
well accommodate the corresponding differential merger rate densities
estimated from GWTC-1 and GWTC-2 (Fig.~\ref{fig:diffrate}; Sec.~\ref{sec.diffr};
Fig.~\ref{fig:diffrate2}; Sec.~\ref{discuss}). The large difference between
the reference $\rate$ and the pessimistic $\rpess$ (and, hence,
between $d\rate/dX$ and $d\rpess/dX$) arises due to large uncertainties
in the spatial number density of GCs, $\rhogc$, that is used as a `tracer'
to estimate the population of clusters formed, until the
present epoch, within the (comoving)
volume enclosed by the detector horizon at $\zmax$ (Sec.~\ref{ratecalc}; Sec.~\ref{discuss}). 
The difference arises as well due to the uncertainties in the lower mass limit,
$\mgclow$, of the progenitors of present-day GCs (Sec.~\ref{ratecalc}; Sec.~\ref{discuss}).
The Model Universe $\rate$ also depends on the power law index, $\alpha$,
and on the lower mass cutoff, $\mcllow$, of the cluster birth mass function
(Fig.~\ref{fig:rzmax}; Sec.~\ref{discuss}). With improving constraints
on (differential) merger rate density from further observations of compact binary merger events,
such widely debated quantities related to large scale structure formation and
cosmic star formation can be better constrained. 
The Model Universe yields
eccentric LVK mergers from YMCs and OCs at the current epoch, with a (reference) merger rate
density of $\recc\approx5.0\peryg$ for $\zmax=1$ (Fig.~\ref{fig:drde}; Sec.~\ref{sec.diffr}).

The Model Universe $\rate$ depends on $\zmax$, maximizing to
$\rate=64.1\peryg$ at $\zmax\approx3.5$, most of the growth
in $\rate$ occurring within $\zmax\lesssim2.0$ (Fig.~\ref{fig:rzmax}; Table~\ref{tab_rates}; Sec.~\ref{sec.rz}).
This $\zmax$ dependence is due to the cosmic variation of star formation rate (the SFH)
and the distribution of merger delay time, $\tmrg$, of the Model Universe
(Fig.~\ref{fig:cosmorate}, right panel), resulting in an inherent
dependence, $\rz(\zevnt)$, of merger rate density on merger-event redshift
(the cosmic merger rate density function; Fig.~\ref{fig:cosmorate}, left panel; Sec.~\ref{sec.rz}).

The Model Universe (differential) merger rate density will further improve
as the computed model grid gets more complete, homogeneous, and extended.
This is demonstrated in Sec.~\ref{discuss} using a model grid (Table~\ref{tab_nbrun}) that
is richer in its high mass end than that of Paper II.
Additional
N-body computations to that end are ongoing and the updated results will be
presented in a future paper. It would also be worth comparing the
outcomes of the present Model Universe with those from Universes
that incorporate alternative cosmic metallicity evolution, \eg, those
of \citet{Rafelski_2012,Madau_2017}. For a more complete and consistent treatment of the formation
and evolution of star cluster population in the Universe over cosmic time
(and, hence, of dynamical merger rate estimates from them),
in relation to the present-day GC population, it is necessary to
combine evolutionary models of clusters as in here (of $\sim10^4\Ms-10^5\Ms$)
with those of lower mass clusters as in, \eg, \citet{Rastello_2019,DiCarlo_2019,Kumamoto_2019}
and of much higher mass GC-progenitor clusters as in, \eg, \citet{Askar_2016,Kremer_2020}.
GWTC-2 has been released whilst preparing this manuscript which is why
the GWTC-2 data is addressed somewhat briefly here and a more
elaborate comparison (as in, \eg, Paper II, which addresses the GWTC-1 merger-event data)
would be worth undertaking. The present work motivates such future lines of research.

This study suggests that with reasonable astrophysical
inputs based on Local Universe and cosmological observations
(Sec.~\ref{ratecalc}; Sec.~\ref{discuss}), dynamical interactions in YMCs and OCs
can, in principle, explain the BBH merger rate density and
the corresponding differential rate densities as estimated, so far, from LVK
GW events, without invoking additional channels for producing compact binary mergers.
As the (differential) merger rate density gets increasingly constrained
with forthcoming merger-event detections, the relative role of
the various channels of compact binary mergers will be better understood.
The present results do not imply that there is only one channel
responsible for the observed properties and rates of GR mergers \citep[\eg,][]{Zevin_2020}.

\section*{Acknowledgements}

SB acknowledges the support from the Deutsche Forschungsgemeinschaft (DFG; German Research Foundation)
through the individual research grant ``The dynamics of stellar-mass black holes in
dense stellar systems and their role in gravitational-wave generation'' (BA 4281/6-1; PI: S. Banerjee).
SB acknowledges the generous support and efficient system maintenance of the
computing team at the AIfA and HISKP. SB acknowledges the pleasant hospitality
and energetic discussions during the Cluster Dynamics Workshop held at
the CIERA, Northwestern University, U.S.A., in December 2018, which has generated
motivation for this work.

\section*{Data availability}

The GWTC-1 data utilized in this article is publicly available at the URL
\url{https://dcc.ligo.org/LIGO-P1800324/public} and is described in the paper
\citet{Abbott_GWTC1_prop}.
The GWTC-2 data utilized in this article is publicly available at the URL
\url{https://dcc.ligo.org/LIGO-P2000434/public} and is described in the paper
\citet{Abbott_GWTC2_prop}.
The redshift-metallicity relation data is obtained from the
public repository provided in \citet{Chruslinska_2019}. Further details
on how these data are accessed are provided in the text.
The simulation data underlying this article will be shared upon reasonable
request to the corresponding author.

%
 
\bibliographystyle{mnras}
\bibliography{bibliography/biblio.bib}

\label{lastpage}

\appendix

\section{An amended set of star cluster models}\label{newset}

\onecolumn
\renewcommand*{\arraystretch}{1.5}
\begin{longtable}{>{\stepcounter{rowno}\therowno}rccclcllrcc}
	\caption[Summary of model calculations utilized in Sec.~\ref{discuss}]
	{Summary of direct N-body evolutionary models of star clusters and their GR-merger yields
	that are used in the merger rate estimates in  Sec.~\ref{discuss}. The columns from left to right give the model
	cluster's (a) ID number, (b) initial mass, $\mcl$, (c) initial half-mass radius, $\rh$,
	(d) metallicity, $Z$, (e) initial fraction of primordial binaries, $\fbin$,
	(f) model evolutionary time, $\tevol$, (g) remnant-mass and PPSN/PSN model,
	(h) remnant natal kick model, (i) BH natal spin model,
	(j) number of GR mergers within the cluster, $\nmrgin$, (k) number of GR mergers
	after getting ejected from the cluster, $\nmrgout$. See Paper II and Table~C1 thereof
	for further details.
	}\label{tab_nbrun}\\
	\hline
	\hline
	\multicolumn{1}{r}{No.} & \mcl/\Ms     & \rh/pc & $Z$ & \fbin & \tevol/Gyr & remnant model & SN kick & BH spin & \nmrgin & \nmrgout \\
	\hline
	\endfirsthead
        
	\multicolumn{7}{c}%
        {{\bfseries \tablename\ \thetable{} -- continued from previous page}} \\
        \hline
	\hline
	\multicolumn{1}{r}{No.} & \mcl(0)/\Ms     & \rh(0)/pc & $Z$ & \fbin(0) & \tevol/Gyr & remnant model & SN kick & BH spin & \nmrgin & \nmrgout \\
	\hline
	\endhead

	\hline \multicolumn{10}{r}{{Continued on next page}} \\ \hline
        \endfoot

        \hline \hline
        \endlastfoot

    &   $2.0\times10^4$ & 2.0       & 0.0002 &  0.10\footnote{The binary fraction is defined
	as $\fbin=2\nbin/N$, $\nbin$ being the total number of binaries and $N$ being
	the total number of members.}
	                                                &  10.9    &   rapid+B16  &  mom. cons.\footnote{
								mom. cons. $\Rightarrow$ momentum conserving natal kick model,
							col. asym. $\Rightarrow$ collapse-asymmetry-driven natal kick model.}
					                                                        &  FM19\footnote{Geneva (MESA)
			                                $\Rightarrow$ BH natal spin model, as in \citet{Belczynski_2020}, based on
							fast-rotating Geneva (MESA) stellar-evolutionary models,
							FM19 $\Rightarrow$ zero natal spin for all BHs \citep{Fuller_2019a}.}
							                                                   &  1    & 0     \\
    &   $2.0\times10^4$ & 2.0       & 0.001  &  0.10    &  11.0    &   rapid+B16  &  mom. cons. &  Geneva  &  1    & 0     \\
    &   $2.0\times10^4$ & 2.0       & 0.001  &  0.10    &  8.7     &   rapid+B16  &  col. asym. &  Geneva  &  1    & 0     \\
    &   $2.0\times10^4$ & 2.0       & 0.005  &  0.10    &  8.8     &   rapid+B16  &  mom. cons. &  FM19    &  1    & 0     \\
    &   $2.0\times10^4$ & 2.0       & 0.01   &  0.10    &  4.4     &   rapid+B16  &  col. asym. &  Geneva  &  0    & 0     \\
    &   $2.0\times10^4$ & 2.0       & 0.01   &  0.10    &  5.8     &   rapid+B16  &  mom. cons. &  Geneva  &  1    & 0     \\
    &   $2.0\times10^4$ & 2.0       & 0.02   &  0.10    &  4.4     &   rapid+B16  &  mom. cons. &  Geneva  &  0    & 0     \\
    &   $3.0\times10^4$ & 1.0       & 0.0002 &  0.00    &  11.0    &   rapid+B16  &  mom. cons. &  Geneva  &  3    & 0     \\
    &   $3.0\times10^4$ & 1.0       & 0.01   &  0.00    &  7.2     &   rapid+B16  &  mom. cons. &  Geneva  &  2    & 0     \\
    &   $3.0\times10^4$ & 1.0       & 0.01   &  0.00    &  11.0    &   rapid+B16  &  col. asym. &  Geneva  &  1    & 1     \\
    &   $3.0\times10^4$ & 1.0       & 0.01   &  0.00    &  10.9    & delayed+B16  &  col. asym. &  MESA    &  4    & 0     \\
    &   $3.0\times10^4$ & 1.0       & 0.02   &  0.00    &  11.0    & delayed+B16  &  col. asym. &  MESA    &  0    & 0     \\
    &   $3.0\times10^4$ & 1.0       & 0.02   &  0.00    &  8.2     &   rapid+B16  &  mom. cons. &  Geneva  &  0    & 0     \\
    &   $3.0\times10^4$ & 1.0       & 0.02   &  0.00    &  7.0     &   rapid+B16  &  col. asym. &  Geneva  &  1    & 0     \\
    &   $3.0\times10^4$ & 2.0       & 0.0002 &  0.00    &  11.0    &   rapid+B16  &  mom. cons. &  Geneva  &  1    & 0     \\
    &   $3.0\times10^4$ & 2.0       & 0.01   &  0.00    &  9.7     &   rapid+B16  &  mom. cons. &  Geneva  &  2    & 0     \\
    &   $3.0\times10^4$ & 2.0       & 0.01   &  0.00    &  11.0    &   rapid+B16  &  col. asym. &  Geneva  &  2    & 0     \\
    &   $3.0\times10^4$ & 2.0       & 0.02   &  0.00    &  11.0    &   rapid+B16  &  col. asym. &  Geneva  &  0    & 0     \\
    &   $3.0\times10^4$ & 2.0       & 0.001  &  0.10    &  11.0    &   rapid+B16  &  mom. cons. &  Geneva  &  2    & 0     \\
    &   $3.0\times10^4$ & 2.0       & 0.001  &  0.10    &  11.0    &   rapid+B16  &  col. asym. &  Geneva  &  0    & 0     \\
    &   $3.0\times10^4$ & 2.0       & 0.01   &  0.10    &  11.0    &   rapid+B16  &  mom. cons. &  Geneva  &  0    & 0     \\
    &   $3.0\times10^4$ & 2.0       & 0.01   &  0.10    &  11.0    &   rapid+B16  &  col. asym. &  Geneva  &  1    & 0     \\
    &   $3.0\times10^4$ & 2.0       & 0.02   &  0.10    &  3.0     &   rapid+B16  &  mom. cons. &  Geneva  &  0    & 0     \\
    &   $3.0\times10^4$ & 2.0       & 0.001  &  0.10\footnote{$\mcrit=5.0\Ms$}
	                                                &  11.0    &   rapid+weak &  mom. cons. &  MESA    &  1    & 2     \\
    &   $3.0\times10^4$ & 2.0       & 0.001  &  0.10\footnote{$\mcrit=5.0\Ms$}
							&  9.3     &   rapid+weak\footnote{30\% of the full B10 wind is applied.}
							                          &  mom. cons. &  MESA    &  0    & 1     \\
    &   $5.0\times10^4$ & 2.0       & 0.0002 &  0.00    &  11.0    &   rapid+B16  &  mom. cons. &  Geneva  &  0    & 1     \\
    &   $5.0\times10^4$ & 2.0       & 0.0002 &  0.00    &  11.0    &   rapid+weak &  mom. cons. &  MESA    &  1    & 0     \\
    &   $5.0\times10^4$ & 2.0       & 0.001  &  0.00    &  11.0    &   rapid+B16  &  mom. cons. &  MESA    &  2    & 0     \\
    &   $5.0\times10^4$ & 2.0       & 0.001  &  0.00    &  11.0    &   rapid+weak &  mom. cons. &  MESA    &  0    & 0     \\
    &   $5.0\times10^4$ & 2.0       & 0.001  &  0.00    &  11.0    &   rapid+B16  &  col. asym. &  MESA    &  2    & 0     \\
    &   $5.0\times10^4$ & 2.0       & 0.005  &  0.00    &  11.0    &   rapid+weak &  mom. cons. &  MESA    &  1    & 0     \\
    &   $5.0\times10^4$ & 2.0       & 0.005  &  0.00    &  11.0    & delayed+B16  &  col. asym. &  MESA    &  4    & 0     \\
    &   $5.0\times10^4$ & 2.0       & 0.01   &  0.00    &  11.0    &   rapid+B16  &  mom. cons. &  Geneva  &  1    & 0     \\
    &   $5.0\times10^4$ & 2.0       & 0.01   &  0.00    &  11.0    &   rapid+B16  &  col. asym. &  Geneva  &  3    & 0     \\
    &   $5.0\times10^4$ & 2.0       & 0.02   &  0.00    &  9.9     &   rapid+B16  &  mom. cons. &  Geneva  &  0    & 0     \\
    &   $5.0\times10^4$ & 2.0       & 0.02   &  0.00    &  11.0    & delayed+B16  &  col. asym. &  MESA    &  3    & 0     \\
    &   $5.0\times10^4$ & 2.0       & 0.0001 &  0.05    &  11.0    &   rapid+B16  &  mom. cons. &  Geneva  &  1    & 1     \\
    &   $5.0\times10^4$ & 2.0       & 0.001  &  0.05    &  10.0    &   rapid+B16  &  mom. cons. &  Geneva  &  4    & 2     \\
    &   $5.0\times10^4$ & 2.0       & 0.001  &  0.05\footnote{$\ftz=0.70$, $\fmrg=0.3$}
                                                        &  11.0    &   rapid+weak &  mom. cons. &  MESA    &  1    & 2     \\
    &   $5.0\times10^4$ & 2.0       & 0.0001 &  0.05\footnote{$\fmrg=0.2$}
                                                        &  11.0    &   rapid+B16  &  mom. cons. &  MESA    &  2    & 2     \\
    &   $5.0\times10^4$ & 2.0       & 0.01   &  0.05\footnote{$\ftz=0.90$, $\fmrg=0.2$}
                                                        &  11.0    & delayed+B16  &  col. asym. &  MESA    &  2    & 0     \\
    &   $5.0\times10^4$ & 1.0       & 0.001  &  0.00    &  11.0    &   rapid+B16  &  mom. cons. &  Geneva  &  1    & 2     \\
    &   $5.0\times10^4$ & 1.0       & 0.001  &  0.00    &  11.0    & delayed+B16  &  col. asym. &  MESA    &  3    & 0     \\
    &   $5.0\times10^4$ & 1.0       & 0.001  &  0.00    &  11.0    & delayed+B16  &  mom. cons. &  FM19    &  5    & 0     \\
    &   $5.0\times10^4$ & 1.0       & 0.01   &  0.00    &  11.0    & delayed+B16  &  col. asym. &  MESA    &  7    & 0     \\
    &   $5.0\times10^4$ & 1.0       & 0.02   &  0.00    &  11.0    & delayed+B16  &  col. asym. &  MESA    &  4    & 4     \\
    &   $7.5\times10^4$ & 2.0       & 0.001  &  0.00    &  11.0    &   rapid+B16  &  mom. cons. &  Geneva  &  2    & 0     \\
    &   $7.5\times10^4$ & 2.0       & 0.001  &  0.00    &  11.0    &   rapid+weak &  mom. cons. &  MESA    &  3    & 0     \\
    &   $7.5\times10^4$ & 2.0       & 0.001  &  0.00    &  11.0    &   rapid+B16  &  col. asym. &  Geneva  &  3    & 0     \\
    &   $7.5\times10^4$ & 2.0       & 0.005  &  0.00    &  11.0    & delayed+B16  &  col. asym. &  MESA    &  4    & 0     \\
    &   $7.5\times10^4$ & 2.0       & 0.01   &  0.00    &  11.0    &   rapid+B16  &  mom. cons. &  Geneva  &  6    & 0     \\
    &   $7.5\times10^4$ & 2.0       & 0.02   &  0.00    &  11.0    &   rapid+B16  &  mom. cons. &  Geneva  &  6    & 0     \\
    &   $7.5\times10^4$ & 2.0       & 0.02   &  0.00    &  11.0    & delayed+B16  &  col. asym. &  MESA    &  3    & 1     \\
    &   $7.5\times10^4$ & 2.0       & 0.0001 &  0.05\footnote{$\fmrg=0.2$}
                                                        &  11.0    &   rapid+B16  &  mom. cons. &  MESA    &  1    & 2     \\
    &   $7.5\times10^4$ & 2.0       & 0.001  &  0.05\footnote{$\ftz=0.95$, $\fmrg=0.2$}
                                                        &  11.0    &   rapid+B16  &  mom. cons. &  MESA    &  1    & 4     \\
    &   $7.5\times10^4$ & 2.0       & 0.001  &  0.05\footnote{$\ftz=0.95$, $\fmrg=0.2$}
                                                        &   9.8    &   rapid+B16  &  mom. cons. &  FM19    &  5    & 1     \\
    &   $7.5\times10^4$ & 2.0       & 0.01   &  0.05\footnote{$\ftz=0.95$, $\fmrg=0.2$}
                                                        &  11.0    &   rapid+B16  &  mon. cons. &  FM19    &  1    & 1     \\
    &   $7.5\times10^4$ & 2.0       & 0.02   &  0.05\footnote{$\fmrg=0.2$}
                                                        &  11.0    & delayed+B16  &  col. asym. &  MESA    &  2    & 0     \\
    &   $1.0\times10^5$ & 2.0       & 0.001  &  0.00    &  11.0    & delayed+B16  &  col. asym. &  Geneva  &  5    & 1     \\
    &   $1.0\times10^5$ & 2.0       & 0.005  &  0.00    &  11.0    &   rapid+B16  &  mom. cons. &  FM19    &  6    & 0     \\
    &   $1.0\times10^5$ & 2.0       & 0.01   &  0.00    &  11.0    &   rapid+B16  &  mom. cons. &  FM19    &  6    & 3     \\
    &   $1.0\times10^5$ & 2.0       & 0.02   &  0.00    &  11.0    &   rapid+B16  &  mom. cons. &  FM19    &  6    & 0     \\
    &   $1.0\times10^5$ & 1.5       & 0.001  &  0.05\footnote{$\ftz=0.95$, $\fmrg=0.2$}
							&  11.0    &   rapid+B16  &  mom. cons. &  FM19    &  8    & 0     \\
    &   $1.0\times10^5$ & 2.0       & 0.001  &  0.05\footnote{$\ftz=0.95$, $\fmrg=0.2$}
                                                        &  11.0    &   rapid+B16  &  mom. cons. &  FM19    &  4    & 0     \\
\end{longtable}
\twocolumn

%
\end{document}